\documentclass{aastex}
\usepackage{spr-astr-addons}
\usepackage{url}
\usepackage{graphicx}
\usepackage{epstopdf}
\RequirePackage{color}

\begin{document}

\title{Polarization Sounding of Pulsar Magnetosphere\\ (Part I)}
\slugcomment{Not to appear in Nonlearned J., 45.}
%% Running heads
\shorttitle{Short article title}
\shortauthors{Autors et al.}

\author{Ulyanov O.M. - oulyanov@rian.kharkov.ua} %\altaffilmark{1}}
 \and \author{Shevtsova A.I. - ashevtsova@rian.kharkov.ua} %\altaffilmark{1}}
%\and
\author{Skoryk A.O. - skoryk.a@rian.kharkov.ua} %\altaffilmark{1}}
\affil{Institute of Radio Astronomy of the National Academy of Sciences of Ukraine (IRA NASU), Chervonopraporna St. 4, Kharkov, 61002, Ukraine}
%\affil{Affilation of Third Author}
%\email{\emaila oulyanov@rian.kharkov.ua}

%\altaffiltext{1}{Institute of Radio Astronomy of NAS of Ukraine}
%\altaffiltext{2}{Second Alternate Affilation.}
%\altaffiltext{3}{Third Alternate Affilation.}

\begin{abstract}
In a cycle of papers, starting with the present one, a fundamental possibility of the polarization sounding of a pulsar magnetosphere using pulsar self-radiation as a test signal will be considered. The main idea of such a sounding is based on the fact that in some models of pulsar magnetosphere the emission frequency depends on the height of its origin above the pulsar surface. For this purpose it is needed to study the dependence of such a parameter as the rotation measure on frequency and pulse phase. We expect that it is possible to register the rotation measure dependence on the central observing frequency and/or the pulse phase during the observations at widely spaced frequencies in the same conditions. The reliable registration of this dependence will result in resolving of the pulsar magnetosphere in depth. It is preferably to carry out the polarization researches in the observational mode of the individual pulses.

In this part of the work the model of polarized pulsed radio emission and the model of weakly anisotropic propagation medium are considered.

In future articles we will show how to determine the polarization parameters of pulsar radiation with the highest precision in presence of different types of recorders on the receiving side. The specifics of polarization observations at decameter wavelengths will be considered. The algorithms for solving the inverse problem for different types of receivers and the variable conductivity of the underlying surface will be developed.
\end{abstract}


%\section*{}
%\label{sec:intro}

\section{Introduction}

Radio pulses observed from a pulsar are generated in its magnetosphere. This region contains strong magnetic fields and the electron-positron plasma \citep{1969ApJ...157..869G,1971ApJ...164..529S}. The electromagnetic processes that occurs in a pulsar magnetosphere lead to the conversion of energy into the energy of the star rotation, the magnetic dipole radiation energy, the current loss or a combination of those conversion of energy of the star rotation into the magnetic dipole radiation energy, the current loss or a combination of those \citep{1977puls.book.....S,1977puls.book.....M,1988Ap&SS.146..205B,2009ARep...53.1146B}.

Since it is impossible to recreate the conditions that occur in the pulsar magnetosphere plasma in a laboratory on the Earth, observations of pulsar radio emission (PRE) in all frequency ranges are of great interest. Studying the properties of the radio-emitting region is of paramount importance for understanding of both the nature of the PRE mechanism and the internal structure of the pulsar magnetosphere. The pulsed nature of radiation in the observer reference frame makes pulsars the good probes to study not only the interstellar medium, but also the pulsar magnetosphere and its pulsar wind.

PRE has a number of features, among which the most important for probing the propagation medium are a broadband spectrum, periodicity of pulsed signals and polarization of its radio emission \citep{1977puls.book.....S,1977puls.book.....M}. The PRE spectrum extends from decameter to the millimeter range \citep{1978SvA....22..588B}, the PRE is coherent and periodic (periods are in the range from milliseconds to seconds), and the degree of linear polarization of giant and anomalously intense pulses is close to 100 \% \citep{2005AJ....129.1993M, 2006ARep...50..562P,atnfCite2014}.

These three factors make it possible to use the most intense pulses of pulsars to probe the propagation medium (including the magnetosphere) without knowing reliably the nature of the coherent radiation of these pulses.

The purpose of this part of the work will be (i) to build a model of polarized pulsed radiation, the models of the propagation medium, (ii) to develop algorithms for determination the dynamically changing polarization parameters of pulse signals and (iii) to demonstrate the capabilities of resolution of pulsar magnetosphere upper layers in depth.

To achieve this goal we need to formulate the direct problem, i.e. to build a model of PRE polarized pulse and to build a model of weakly anisotropic propagation medium of the polarized radiation. Next, it is needed to develop an algorithm for estimating the propagation medium parameter such as the $RM$ (Rotation Measure). After that the inverse problem can be solved and we can estimate the polarization parameters of pulsed radiation in the reference frame associated with the pulsar.

All the modeling is carried out using numerical methods. The properties of the polarized radiation associated with occurrence of the Faraday effect in the propagation medium were used for the estimation of polarization parameters. The effect is manifested in rotation of the polarization plane  of the elliptically polarized radio emission (in general case), while it propagates along the line of sight. This assumes the presence of parallel to line of sight component of the magnetic field
$ \vec{B_{\parallel}} = \vec{B} \cdot \vec{k} (\omega) / \vert \vec{k}(\omega) \vert $, where $ \vec{B} $ is the magnetic induction vector, $ \vec{k} $ is the wave vector, $ \omega $  is the cyclic frequency.

 The four Stokes parameters $ {I(\omega_c,\psi )} $, $ {Q(\omega_c,\psi )} $, $ {U(\omega_c,\psi )} $ , $ {V(\omega_c,\psi )} $ will be considered as the polarization parameters. Relative polarization parameters are: the degree of polarization $ {P_t(\omega_c,\psi )} $ , the degree of linear polarization $ {P_l(\omega_c,\psi)} $,  the degree of circular polarization $ {P_c(\omega_c,\psi )} $, the degree of depolarization $ {P_d(\omega_c,\psi )} $, ellipticity $ {\varepsilon(\omega_c,\psi )} $. Additional parameters that need to be evaluated are the position angle traverse along the pulse profile $ {\chi(\omega_c,\psi )} $ and the dependence of rotation measure on the pulse phase $ {RM(\omega_c,\psi )} $ (also these parameters depend on the central emission frequency $ \omega_c $).

The decameter range is the most difficult for the research due to the various kinds of the interference of natural and artificial origin. All the known propagation effects are expressed most sharply in this range. They are the dispersive time delay of propagation signal at different frequencies, the scattering in the interstellar medium, refraction, Faraday effect \citep{2010MNRAS.403..569W,2011MNRAS.417.1183W}, etc.

Along with the limitations of this range, there are several positive factors that augur well for the successful solution of the problem. For example, the occurrence of anomalously intensive pulses of PSR B0809 +74, PSR B0943 +10, PSR B0950 +08, PSR B1133 +16, \citep{2012ARep...56..417U} and the giant pulses of the pulsar in the Crab Nebula PSRB 0531 +21 \citep{2006ARep...50..562P}.

Considering the complexities of the polarization researches in this range, there are two circumstances that allow to hope for obtaining the polarization parameters of the PRE in the observations with the radio telescopes UTR-2 with linearly polarized dipoles \citep{MegnA.V.1978, Braude1978} and URAN-2 with crossed dipoles \citep{2005KFNTS...5...43B}. First, the radiation cone of most known pulsars expands with  decreasing frequency \citep{2013MNRAS.431.3624Z}, which makes it possible to detect the pulsar radiation components that are not visible at higher frequencies. Second, the Faraday effect that occurs during the propagation of the PRE in the interstellar medium with a weak regular component of the galactic magnetic field parallel to the line of sight \citep{1970pewp.book.....G, 1987tfia.book.....G, 1997riap.book.....Z, 2013BaltA..22...53U, UlyanovShevtsova2013, UlyanovShevtsova2014} . To date, there have been several polarization observations at the radio telescopes UTR-2 \citep{2013BaltA..22...53U}, URAN-2 \citep{2005KFNTS...5...43B}, DKR 1000 and BSA 100 \citep{ 1974AZh....51..927S, 1983SvA....27..322S, 1988SvA....32..177S, 1989puls.book...42S}, employing this effect. However, in all these works, the Fourier analysis of the polarization response or inscribing the simple harmonic signal (sine/cosine) in the observational data by the method of the minimum mean square deviations (MMSD) were used in the interpretation of the obtained results \citep{1989puls.book...42S}. Alternatively, we propose a more complex model of the polarized signal in this paper.  Application of this model allows to receive all the polarization parameters of the pulsed radiation. The methods developed here could be used with the radio telescopes UTR-2, URAN-2, GEETEE, DKR 1000, BSA, LOFAR, LWA, SKA. Exactly these radio telescopes operate or will operate at the lowest frequencies available for observations from the Earth \citep{2007ApJ...665..618B, 2011A&A...530A..80S, 2013ApJ...768..136E}.

\section{Polarized Pulsar Radio Emission (Model Representation)}

Despite more than forty five years of PRE researches \citep{1968Natur.217..709H, 1968Natur.218..126P, 1969ApL.....3..225R, 2007ApJ...665..618B, 2011A&A...530A..80S, 2013ApJ...768..136E}, to date the mechanism of coherent radio emission remains unclear. From the observational data obtained in different frequency ranges, it is clear that the radio emission of pulsars is polarized \citep{1968Natur.217..709H, 1968Natur.218..126P, 1969ApL.....3..225R, 2003ASPC..302..179M, 2006A&A...448.1139S, 2009ApJ...692..459M}. Typically, the linear polarization dominates \citep{ 2002ARep...46..309S, 2003ASPC..302..179M}. However, the component with circular polarization is also present \citep{ 2003ASPC..302..179M, 2009ApJ...692..459M}.

On the other hand, in the phenomenological models of Rankin \citep{1983ApJ...274..359R, 1983ApJ...274..333R, 2002ARep...46..309S, 2006A&A...448.1139S} with observational confirmation, the so-called "core" and "cone" components have different polarization properties. In the central («core») radiation component the degree of circular polarization is increased. In the side («cone») radiation components the degree of linear polarization is increased.

For a better understanding of the processes associated with the polarization transformation of radio emission at low frequencies during its propagation, we used numerical simulations of  polarized pulses and of the propagation medium \citep{1997riap.book.....Z, 2013BaltA..22...53U, UlyanovShevtsova2013, UlyanovShevtsova2014}. In this section we present a model of polarized radio pulse of a hypothetical pulsar. The proposed model allows to extend the analysis bandwidth to the values actually used for the observations in the decameter frequency range.

The short radio pulses that have the certain repetition period, corresponding to the pulsar period $\text{P}_{PSR}$ were modeled. Value of $\text{P}_{PSR}$ = 1 second was chosen as a typical value of this parameter for ordinary (non-millisecond) radio pulsars. The elliptical polarization was used in the model representations of the PRE in order to simulate the presence of both linear and circular polarizations in radiation.

The model of polarized radio pulse was simulated in accordance with the following items:

\paragraph{1. Generating the grid of carrier frequencies.}

\begin{figure}[h]
%\imagei
\centering
\includegraphics[width=70mm]{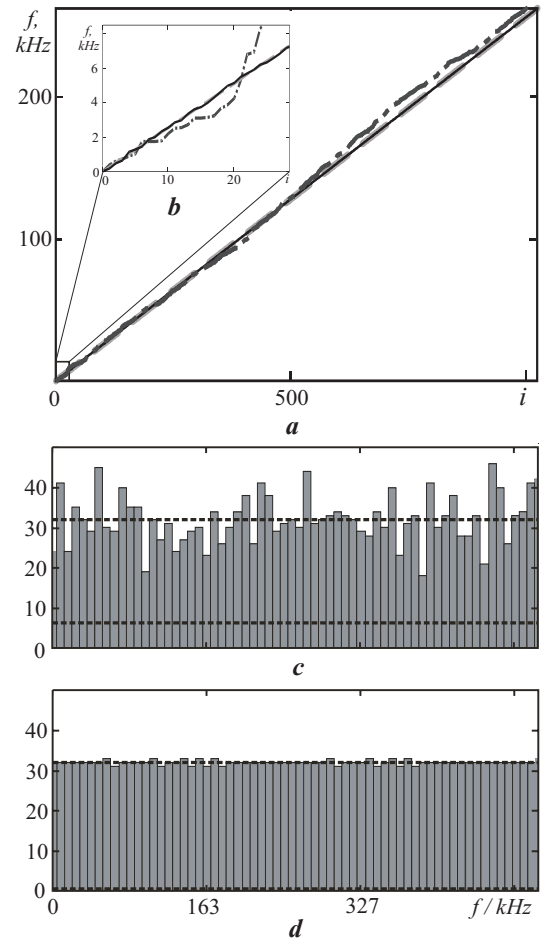} %
\caption{%
Frequency distribution in the receiving band. a) is the frequency distribution versus indices, b) is the enlarged fragment at the beginning of the frequency distribution range; c) is  a histogram of the frequency distribution in the entire reception band in the case of global uniform distribution; d) is the same as in c) for a uniform distribution of the local frequencies.}
\label{fig1}
 %% no full stop at the end
\end{figure}

At first, the carrier frequencies $f_i$ were generated using a random number generator based on uniformly distributed random variables (subscript hereafter emphasizes the discrete nature of this model and will be omitted below in the integral, differential and matrix equations). These frequencies are evenly distributed throughout the reception band. Also, the two methods of generation were used. The first method provides random, uniform distribution of each spectral interval that is not resolved by further discrete Fourier analysis. The second method generates a uniform distribution of frequencies across the reception band. The difference between the first and the second method is that in the former case both the global and the local uniformity of the entire frequency range filling is maintained. In the latter case, only the global uniformity fill throughout the operating band was maintained, and the local frequency areas can have the uneven filling. In both cases, the obtained set of frequencies was sorted in ascending order. The sets of frequencies, obtained with such methods, are used in the future as carrier frequencies of the model signals (see Fig. \ref{fig1}).

As a central reception frequency was selected $F_c = 23.7$ MHz. This frequency range is used for observations with radio telescopes UTR-2  \citep{UlyanovZaharenko2006,  UlyanovDeshpande2007, 2008ARep...52..917U, 2012OAP....25...35U} and URAN-2 \citep{2005KFNTS...5...43B}. The registration band $\bigtriangleup F$ is chosen relatively small $\bigtriangleup F \approx 0.5$ MHz. For the adopted model, as it will be shown later, this restriction is not of essential character;

\paragraph{2. Generating amplitude for a given frequency grid.}
Two arrays of amplitudes were formed for polarization channels A and B respectively. The presence of a subscript indicates that all simulations were carried out using numerical methods (hereafter the subscript in the equations will be omitted). Each amplitude in the array matches its carrier frequency and longitude. Channels A and B represent two orthogonal linear polarization components that form the polarization ellipse. It is assumed that the local reference frame, where the elliptical polarization exists, is associated with the central frequency $f_i$ of the radiation, the pulsar longitude $\psi_i$ and the local magnetic field $\vec{B}(f_i, \psi_j)$ at a given point of the pulsar magnetosphere.

First, the initial arrays were formed by a random number generator. These arrays have a Rayleigh distribution and the specified standard deviations $ \sigma_1 $ and $ \sigma_2 $: $ A_0(f_i, \psi_j)=\sqrt {-2 {\sigma_{1}}^2 \ln(1 - x_{i,j})} $ and $ B_0(f_i, \psi_j)=\sqrt {-2 {\sigma _{2}}^2 \ln(1-y_{i,j})} $ where $ x_{i,j}$ and $y_{i,j}$ are independent random variables that are uniformly distributed on the interval [0, 1]. Thus, the ratio of mean square deviations $\sigma_1/\sigma_2$ corresponds to the ratio of amplitudes of the expectations in these channels and determines the ellipticity.

To provide a correlation between the values of the amplitudes in adjacent frequencies, the following operations were carried out. Original array $ A_0(f_i, \psi_j) $ was transformed to an array $  A(f_i,\psi_j) $ using a fast Fourier transform (FFT), then a low pass filter (LPF) was applied  and the inverse FFT of the transformed array to the original domain was performed. Similar actions were carried out for the array $ B_0(f_i, \psi_j) $. As a result, two sets of correlated amplitudes for the channels A and B was obtained: $ A(f_i,\psi_j)= IFFT[{FFT[A_0(f_i,\psi_j)] \cdot H(\Omega_k )}] $, $ B(f_i,\psi_j)= \\ IFFT[{FFT[B_0(f_i,\psi_j)] \cdot H(\Omega_k)}] $. Here FFT [*] and IFFT [*] are direct and inverse FFT, respectively, $ \Omega $ is the argument in FFT, $ H(\Omega_k ) $ is the transfer characteristic of the LPF. The ellipticity of the simulated radio emission is $ \langle  { \varepsilon(f_c,\psi_j) } \rangle_f  = \langle { B(f_i,\psi_j)/A(f_i,\psi_j)} \rangle_f  $, where $ \langle * \rangle_f $ denotes averaging over frequency, $ f_c $ is the central frequency of the range that is averaged, and $  { B(f_i,\psi_j)\leq A(f_i,\psi_j)} $. The monotonic variation of the longitude was set for the ellipticity, that is described below in the item 6. The created arrays are shown in Fig. \ref{fig2}:
\begin{figure}[h]
%\imagei
%\centering
\includegraphics[width=84mm]{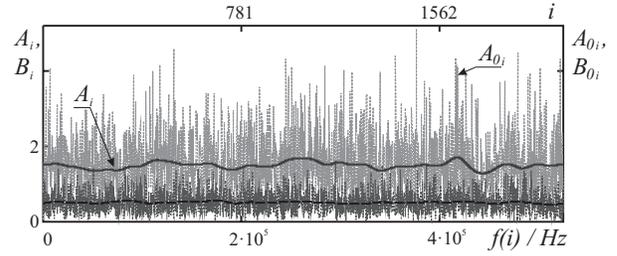}
\caption{%
The set of initial random amplitudes  $ A_0(f_i, \psi_j) $ and $ B_0(f_i, \psi_j) $ distributed over the Rayleigh law and the values $  A(f_i,\psi_j) $ and $  B(f_i,\psi_j) $ converted from it. Gray lines  are the basic amplitude arrays in the polarization channels A and B: $ A_0(f_i, \psi_j) $ and $ B_0(f_i, \psi_j) $, black solid and black dotted lines are the sets of the correlated amplitudes in the receive band: $  A(f_i,\psi_j) $ and $  B(f_i,\psi_j)$.
}
\label{fig2} %% no full stop at the end
\end{figure}
\paragraph{3. Formation of the signal initial phase.}
The array of the signal initial phases $\varphi_{i,j}$ was created using random number generator with a uniform distribution in the angular range [0, 2$\pi$]. At the same time, the mutual phase between channels A and B at any frequency  was equal to $-\pi/2$. It may be noted that the sign at $\pi/2$ depends on the direction of rotation of the resulting elliptical polarization in the observer reference frame: $+\pi/2$ is the right elliptical polarization, $-\pi/2$  is the left elliptical polarization \citep{Landau1969, 2003ASPC..302..179M}.
\paragraph{4. Pulse formation.}
A set of noise-like carrier frequencies has been modulated by amplitude with an envelope that had the shape of a Gaussian $ G_1(\psi_j, a ,\sigma _{G1} )=\dfrac{1}{\sigma_{G1} \sqrt{2 \pi}}\exp(-{\dfrac{(\psi_j-a)^2} {2 ({\sigma_{G1}})^2}}) $, where $ a $ is a phase corresponding to the pulse center, $ \sigma_{G1} $  is the standard deviation (see Fig. 3a). Thus pulsed nature of the pulsar signal in the observer's reference frame was simulated. The pulse width (the width of the envelope) does not exceed 15\% of its duration at half-intensity and/or 27\% at 10\% intensity. In all the figures the dotted lines indicate the boundaries of the pulse at 10\% intensity.

The noise-like signals propagating as transverse waves along the $ x $-axis are shown in the equations (\ref{eq1}).
For an index $i \leq N\!q$, where $f_{N\!q}$ represents the Nyquist frequency, continuous in time analytic signal is given by the next equation:
\begin{equation}                                                          % Eq 1
\begin{array}{l}
 {\dot E}_{0x} ({\omega_i,\psi_j,z})=A(f_{i},\psi_j) \cdot e^{-i(\vec k(\omega_{i}) \vec z-\omega_i t +\varphi_{i}) } \\
 {\dot E}_{0y} ({\omega_i,\psi,z})=B(f_{i},\psi_j) \cdot e^{-i(\vec k(\omega _{i}) \vec z-\omega_i t +\varphi_{i} - \pi /2)} \\
 {\dot E}_{0z} ({\omega_i,\psi,z})=0
\end{array},
\label{eq1}
\end{equation}
where $\vec k(\omega_{i})$ is a set of wave vectors, $\omega _{i}=2 \pi f_{i}$  is a cyclic frequency, $\psi_j$ is a pulse phase. For  $i > N\!q$  $~{\dot E}_{0x,0y,0z} ({\omega_i,\psi_j,z})=0$. The factor $ e^{i \omega_i t}$  can be excluded from the right side of the system of equations (\ref{eq1}) without loss of generality.
\paragraph{5. Generation of additive white noise.}

For each frequency $f_i$, a set of white noise samples was generated. These samples had a normal distribution with zero average value. Next, the white noise was combined additively with a signal from equation (\ref{eq1}). We used the signal to noise ratio $(S / N) > 100$ for greater clarity of the reported results and more accurate their interpretation.

\paragraph{6. Determination of ellipticity.}

\begin{figure}[h]
    %\imagei
    %\centering
    \includegraphics[width=84mm]{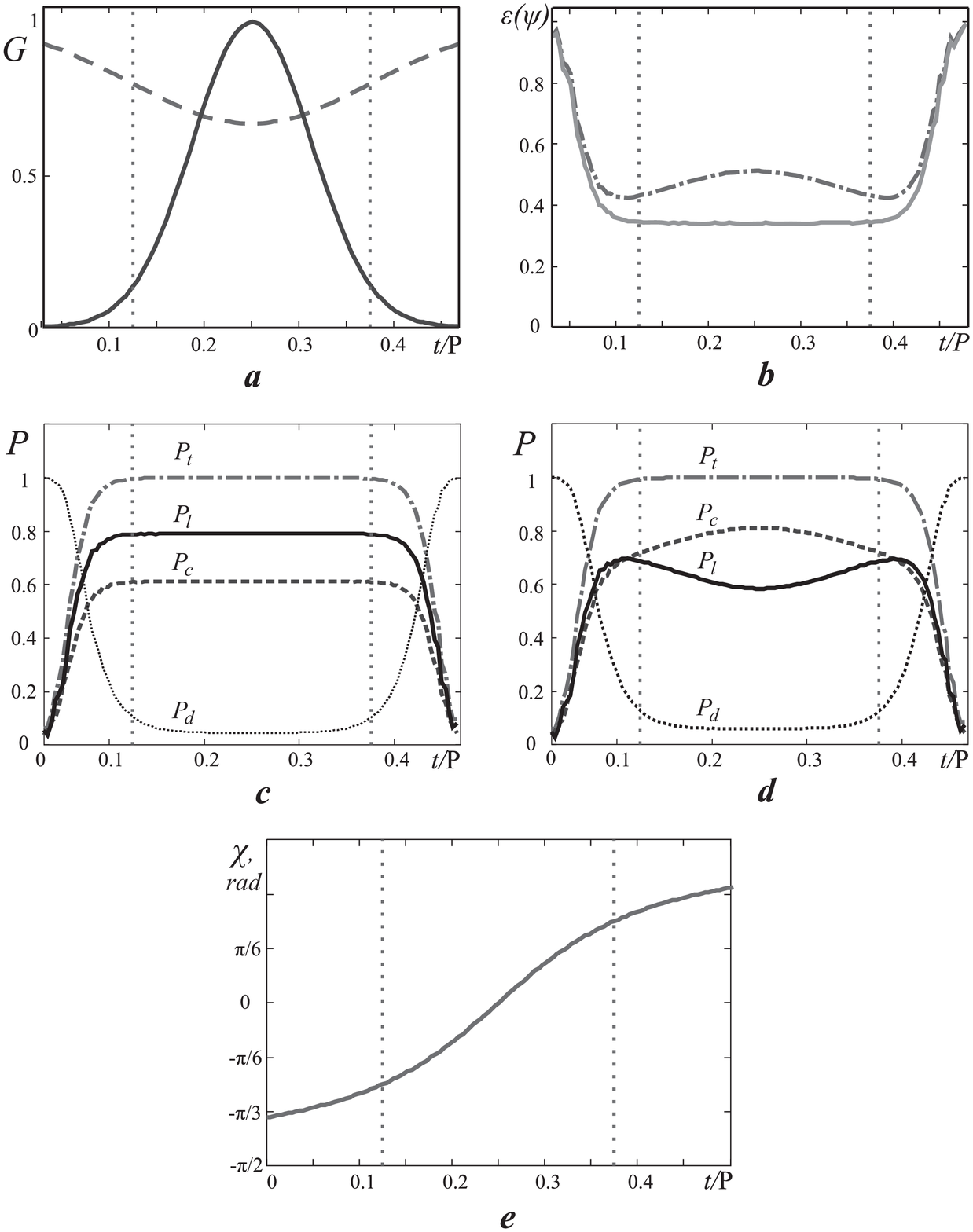}
    \caption{%
        Simulation of the pulse parameters. a) solid line is the shape of average profile pulse envelope $G_{1}(\psi_j, a,\sigma_{G1} )$, dotted line is the function, modulating ellipticity along the average pulse profile $ 1-G_{2}(\psi_j, a,\sigma_{G2} ) $ ; b) the behavior of ellipticity along the pulse profile in the two cases; c) the relative polarization signal parameters $(P_t, P_l, P_c, P_d)$ at a constant ellipticity; d) the polarization degrees in the case of monotonic change of ellipticity along the pulse profile; e) model traverse of the position angle $ {\chi(\omega_c,\psi_j)} $ along the pulse profile;
    }
    \label{fig3} %% no full stop at the end
\end{figure}

For the created pulse in two polarizations the ellipticity $  { \varepsilon(f_i,\psi_j) } $  was specified either  independent of the pulse longitude $ \psi_j $ (as described in the paragraph 2) or monotonically variable  along the pulse profile (Fig. 3b). To change the ellipticity along the pulse profile the channel A was  additionally amplitude modulated by function $ (1-G_2(\psi_j, a ,\sigma _{G1} )) $, shown in Fig. \ref{fig3}$ a $ as a dashed line. The function $G_2$ is similar to $G_1$, but with a different value of the standard deviation $\sigma_{G2}$. Changes in the ellipticity were set symmetrically with respect to the pulse center in such a way: the degree of circular polarization grew to the pulse center and the degree of linear polarization grew towards the pulse edge (see Fig. \ref{fig3}$ c $ and \ref{fig3}$ d $). Because of the noise outside the pulse the ellipticity is identified incorrectly. If the noise power in both polarization channels is identical, away from the pulse one can see a false circular polarization.

Equation (\ref{eq2}) describes the formation of a pulsed signal with a set value of ellipticity along the pulse profile with the additional white noise. The elements $ { \dot E}^0_x(\omega_i,\psi_j,z), {\dot E}^0_y ({\omega_i,\psi_j,z}), {\dot E}^0_z ({\omega_i,\psi_j,z}) $ represent a column vectors of signal components in the local reference frame of the source. We denote this frame of reference by the index "0".
\begin{equation}                                                          % Eq 2
\begin{array}{l}
 {\dot E}^0_x(\omega_i,\psi_j,z)
  =~ G_{1}(\psi_j, a,\sigma_{G1} ) \cdot (1-G_{2}(\psi_j, a,\sigma_{G2} ))\\
  ~~~~~~~~~~~~~\cdot {\dot E}_{0x} ({\omega_i,\psi_j,z}) + N_x( \omega_i,\psi_j) \\
 {\dot E}^0_y ({\omega_i,\psi_j,z})~=~ G_{1}(\psi_j,a,\sigma_{G1})\\
  ~~~~~~~~~~~~~ \cdot {\dot E}_{0y} ({\omega_i,\psi_j,z})+ Ny(\omega_i, \psi_j) \\
 {\dot E}^0_z ({\omega_i,\psi_j,z})~= ~ Nz(\omega_i, \psi)
\end{array},
\label{eq2}
\end{equation}

Let use the term "local reference frame" in the meaning, that is used in the book \cite{1992spai.book.....T}. If there is a model dependence of relative velocity of local reference frame from some parameters, it is expedient to distinguish such local reference frames in analytical sense. As for pulsars radiation the cyclic frequency $ \omega $ and the pulse phase $ \psi $ can be such a parameters, for example $ \vec{V}(\omega,\psi) $. At the same time as analytical describing of local reference frame the physical ability to resolve $ i $-th or $ j $-th local reference frame in conditions of concrete experiment is determining. So if the experimental limit of relative frequency resolution is $ \Delta f/f_c \geqslant 10^{-6} $ then the observer that researches for example the Doppler-effect of some spectral line cannot resolve the two local reference frames the relative velocity $ |\varDelta \vec{V}_{i,j}| / c$ of which is less then that limit.

As were pointed in the item 2 we will associate the local reference frame with the central frequency. This frequency determines the effective altitude above the pulsar surface from which the emission originates. With this the pulse phase defines the azimuthal coordinate of this emission. Below not to make it more complicate we will speak about one local reference frame in coordinates $ (\omega,\psi) $.  At the same time while data processing and interpretation we will try to achive tha maximum available resolution of researching parameters in coordinates $ (\omega_c,\psi) $.

\paragraph{7. Position angle.}

 The position angle  traverse  ($P\!A$) is set along the main pulse longitude. For this the smooth slowly varying along the pulse profile monotonic function $ \chi (\omega_c,\psi_j) = \arctan \left({(\psi_j-a)}/ {m_c} \right)  $ has been used (Fig. \ref{fig3}$ e $). The denominator of the function argument could be changed, simulating different slope of the $ P\!A $ traverse in the pulse window. Similar form of $ P\!A $ traverse actually observed in polarized PRE in different frequency ranges \citep{1969ApL.....3..225R, 1983ApJ...274..359R, 1983ApJ...274..333R,  Malov2004, UlyanovZaharenko2006, 2006MNRAS.365..353K, 2011ARep...55...19M,2013ApJ...768..136E}. One can represent the traverse of position angle in the form of the rotation matrix. Then the signal from the reference frame "$0$" can be translated to a coordinate system associated with the upper pulsar magnetosphere. The mentioned system we denote by the superscript "$a$".

\begin{eqnarray}
 \label{eq3}                                                  %(3)
\left[
\begin{array}{c}
{{\dot E}_{x}}^{a}(\omega_i,\psi_j) \\
{{\dot E}_{y}}^{a}(\omega_i,\psi_j) \\
{{\dot E}_{z}}^{a}(\omega_i,\psi_j)
\end{array}
\right]= ~~~~~~~~~~~~~~~~~~~~~~~~~~~~~~~~~~ \nonumber \\
\left[
\begin{array}{ccc}
\cos{(\chi (\omega_c,\psi))} & -\sin{(\chi (\omega_c,\psi ))} & 0 \\
\sin{(\chi (\omega_c,\psi ))} & \cos{(\chi (\omega_c,\psi))} & 0 \\
0 & 0 & 1
\end{array}
\right] \nonumber\\
 \times \left[
\begin{array}{c}
{{\dot E}_{x}}(\omega_i,\psi_j) \\
{{\dot E}_{y}}(\omega_i,\psi_j) \\
{{\dot E}_{z}}(\omega_i,\psi_j)
\end{array}
\right],
\end{eqnarray}

Thus, two independent polarization channels A and B were formed (see eq. (\ref{eq2}) ). In these channels simulated pulse signals in the reference frame of the radiation source have two orthogonal linear polarization and the specified position angle traverse. These signals match left elliptical polarization \citep{ Landau1969, 2003ASPC..302..179M}. The resulting signals have the specified average ellipticity which depends on the pulse phase and the predetermined $ S/N $ ratio. We assumed that the pulsed radio emission can be formed at different heights from the pulsar surface. These heights depends not only on the frequency of the radiation, but also on the pulse phase. That is, the height depends upon the local magnetic field strength and the local plasma concentration simultaneously. Thus polarized pulsed signals were modeled in the most general representations.

These model representations of the signals are formed on the assumption that the signals are highly correlated in a relatively narrow band of received frequencies. Furthermore, it is assumed here that for each pulse longitude the investigated signals have a fixed ellipticity. This assumption is satisfied at distances from the pulsar surface greater than the critical radius of the polarization \citep{Landau1969, 2006MNRAS.366.1539P, 2006MNRAS.368.1764P, 2010MNRAS.403..569W}.

Then one can proceed to the model representation of the medium, where the polarized pulsed radiation propagates.

\section{Propagation medium (model representation)}

\begin{figure}[h]
    %\imagei
    \includegraphics[width=84mm]{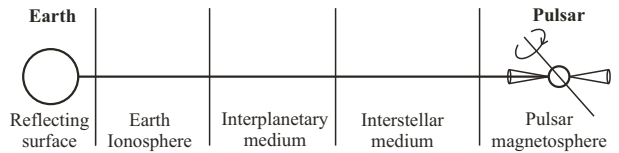}
    \caption{%
        Layered model of the propagation medium.}
    \label{fig4}
\end{figure}

At this stage of research the propagation medium can be represented most conveniently as a set of layers (see Fig. \ref{fig4}). Under the propagation medium we assume weakly anisotropic cold plasma with certain specified parameters.

We distinguish the pulsar magnetosphere, the interstellar medium, the interplanetary medium, the Earth's ionosphere and the reflecting surface. In the pulsar magnetosphere the processes of radiation, formation and propagation of polarized signal occur. Parameters of the following three environments may vary either by known analytical law, or layer can be characterized by homogeneous medium. For example, the average electron concentration in the solar wind varies inversely as the square of the distance from the Sun, and propagation parameters in the interstellar medium (the dispersion measure and the rotation measure) can be considered as the corresponding constants. Finally, in the laboratory system the effects of wave reflection from the underlying surface should be taken into account.

We will consider two main effects that have the greatest influence on propagating signals. They are the frequency dispersion delay of the signals and the rotation of the polarization plane along the line of sight (the Faraday effect) \citep{ 1970pewp.book.....G, 1987tfia.book.....G, 1997riap.book.....Z}. The first effect is explained by the presence of electrons on the propagation path, due to which the speed of the pulses propagation at different frequencies is different. The second effect is connected to the presence of the longitudinal component of the magnetic induction vector along the line of sight. This leads to a difference in refractive indices for the normal modes of the elliptically polarized radiation \citep{1970pewp.book.....G,1997riap.book.....Z}, respectively to different phase/group propagation velocities.

In considered models, we will use the eikonal equation \citep{1987psr..book.....R, 2006MNRAS.365..353K}. This approach allows to represent both the dispersion signal delay in a cold plasma and the Faraday rotation of the polarization plane from a unified position.

For the selected case, the eikonal equation looks as follows: $  \nabla {[\varphi(\omega)]} = n(\omega) \vec k (\omega) $, where $ \nabla[*]  $ is the gradient operator, $ \varphi(\omega) $ is the phase of the corresponding signal harmonic component, $ n(\omega) $ is the refractive index of the medium. This equation is applicable when the nonlinear effects are yet not present in the propagation medium. In a weakly anisotropic medium, the refractive indices for the ordinary wave $ n_{O}(\omega) $ and the extraordinary wave  $ n_{X}(\omega) $  will be different.

\begin{figure}[h]
    %\imagei
    \includegraphics[width=84mm]{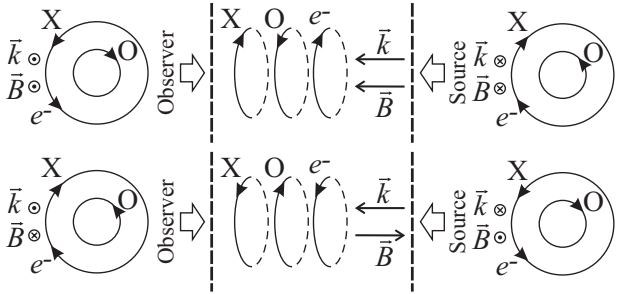}
    \caption{%
        Mutual rotation direction of the ordinary ($O$) and the extraordinary ($X$) waves with respect to the electron ($e^-$) rotation in various configurations of the wave vector $ \vec {k} $ and the magnetic field vector $ \vec{B} $. Vectors directed to the observer are denoted by the point in a circle. Vectors that are directed away from the observer are denoted with a cross in a circle.} %% no full stop at the end
    \label{fig5}
\end{figure}

Next we consider the case of quasi-longitudinal propagation. According to \cite{1970pewp.book.....G} and \cite{1977ewcp.book.....Z, 1997riap.book.....Z}, we define the ordinary wave in the observer reference frame (laboratory reference frame - LRF) as a wave having a right circular polarization (see Fig. \ref{fig5}). The direction of the wave vector coincides with the direction of the magnetic field and the magnetic field itself is directed toward to the observer. Accordingly, the extraordinary wave will have a left-hand circular polarization under the same conditions. If the magnetic field is directed away from the observer to the source, and the wave vector is in the opposite direction, the ordinary wave will have a left-hand circular polarization, and extraordinary - right circular polarization (see Fig. \ref{fig5}).

In that context the left-and right-hand polarizations are defined in the reference frame of the observer. In any reference frame the polarization of the ordinary wave ($O$) has an opposite direction to the direction of the electron rotation in a magnetic field and the polarization of the extraordinary wave ($X$) coincides with it.
Now we define the refractive indices $ n_{O}(\omega) $ and $ n_{X}(\omega) $ in the propagation medium by the following equation:

\begin{equation}
~~~~~~~~~~~~{n_{O,X} (\omega)}= \sqrt{1-{\omega_p^2}/ {\omega(\omega \mp \omega_H)}} \nonumber
\end{equation}
where $ \omega_p = \sqrt{4 \pi e^2  N_e /m_e}  $  is the plasma cyclic frequency, $ e $ is the electron charge, $ N_e $ is the local electron concentration, $ \omega_H  = e |{ \vec{B}}|/{(m_e \gamma  c)}  $ is the gyrotropic cyclic frequency, $ \vec{B} $ is the magnetic induction vector, $ m_e $ is the rest mass of the electron, $ \gamma $ is the electron Lorentz factor (hereinafter in all considered layers except the pulsar magnetosphere, we assume $ \gamma=1 $ ), $ c $ is the speed of light.
Let us consider under which conditions and limits the use of equations of quasi-longitudinal propagation is fair. In general, at presence of weak magnetic field, the following dimensionless parameters are the determining values in the analysis of radio waves normal modes:
$$ u(\omega )=({\omega_{H} /\omega})^2 = (e  |\vec{B}|/(m_e c \omega ))^2 $$
and
$$ v(\omega )=({\omega_{p} /\omega})^2=4 \pi e^2 N_e/(m_e  \omega^2 ) $$

These parameters are different for previously allocated layers of the propagation medium (interstellar medium (ISM), the interplanetary medium (IPM), the Earth's ionosphere (EI)).
To determine the ellipticity of the radio emission normal modes in a cold anisotropic plasma (\cite{1997riap.book.....Z}), we used the following equation :

\begin{gather}                                                                  %eq4
\label{eq4}
M_{O,X}=
\nonumber\\
 \dfrac {2 \sqrt{u}(1-v) cos(\angle \vec{k} \vec{B}) }  {u \! \cdot \!{sin^2\!(\angle \vec{k} \vec{B})} \pm \! \sqrt{u^2 \!\! \cdot \! sin^4({\angle \vec{k} \vec{B}})+ 4 u (1\!-\!v)^2 cos^2({\angle \vec{k} \vec{B}})}}
\end{gather}
where $ M_{O,X} $  is ellipticity ($  { \varepsilon(f_i,\psi_j) } $ see item 6) of the ordinary and extraordinary waves, $ \angle \vec{k} \vec{B} $ is the angle between the wave vector and the magnetic field in the propagation medium (hereinafter referred to the wave vector $ \vec k $ we use the notation, omitting the frequency dependence).

The parameters $u$ and $v$ in the galactic interstellar plasma are defined as follows: $u = 1.39\cdot 10^{-14}$, $v = 4.3\cdot 10^{-9}$. Average value of magnetic field induction vector $ \langle |{{\vec B}_{IMS}| }\rangle \sim  1~ \mu $G. The selected observation frequency $f_c =23.7$ MHz.

\begin{figure}[h]                                                           %fig6
%   \imagei
    \includegraphics[width=84mm]{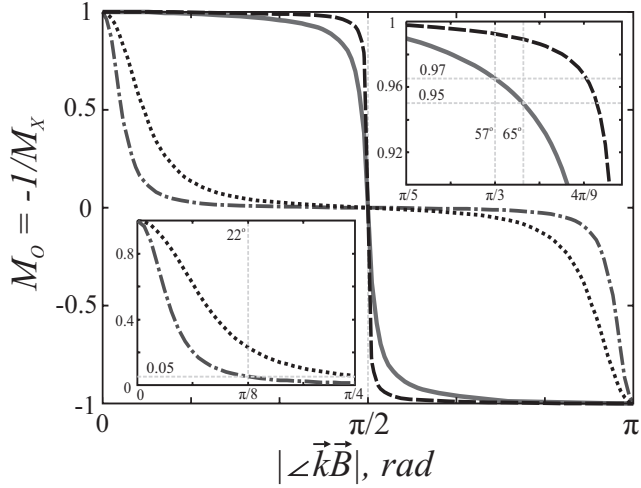}
    \caption{%
    Ellipticity dependencies of the ordinary and the extraordinary waves at frequencies 23.7 MHz and 111 MHz for different values of magnetic induction. The solid line is the ellipticity in the Earth's ionosphere at frequency 23.7 MHz; the dashed line is the ellipticity in the Earth's ionosphere at 111 MHz; the dotted line is the ellipticity in the pulsar magnetosphere at 111 MHz for  $ |\vec B_{PSR}|=1000$ G; the dash-dotted line denotes the ellipticity in the pulsar magnetosphere at the frequency 23.7 MHz ( $|\vec B_{PSR}|=1000$ G).} %% no full stop at the end
    \label{fig6}
\end{figure}

Analysis of the equation (\ref{eq4}) with parameters $u$ and $v$ for the ISM showed that for all angles between and except the angles close to $\pm \pi/2$ the propagation at frequencies above 20 MHz has the quasi-longitudinal character with normal modes of the right and left circular polarization (see Fig. \ref{fig6}).

Interplanetary medium, we present as a layer that is located between the Earth's orbit (1 astronomical unit (AU)) and conventional boundary of the solar system, located at a distance of 50 AU. This choice is motivated by the fact that observations in the decameter range, are generally carried out at high angles of elongation at night. Electron density at the Earth's orbit using satellites data varies from 2 cm$^{-3}$ up to 10 cm$^{-3}$ for the not perturbed solar wind conditions. Magnetic induction of the solar magnetic field on the Earth's orbit reaches an average of 7 nT = 70 $\mu$G. Thus for observation frequency 23.7 MHz coefficients $u$ and $v$ in the interplanetary medium do not exceed $u \leqslant 6.8 \cdot 10^{-11}, v \leqslant 1.43\cdot 10^{-6}$.

Propagation conditions in the Earth ionosphere should be considered in more detail. The magnitude and the direction of the magnetic induction depends on the geographical coordinates of the radio telescope phase center. As in the next part of this work the observational data for the mid-latitude radio telescope UTR-2 will be presented, all of the data from the equation (\ref{eq4}) will be given for its location. UTR-2 is located at the coordinates 49$^{\circ}$38$'$17.6$''$ North latitude and 36$^{\circ}$56$'$28.7$''$ East longitude at an altitude of 180 meters above the sea level. Typical values of the magnetic induction in the UTR-2 phase center have the following parameters $ |{\vec{B}_I}|=50419 \cdot 10^{-9}  $ T, i.e. about 0.5 G. Its vertical and horizontal components, respectively, were: $ |{\vec{B}_{Iv}}|=46311.6 $  nT and $|{{\vec{B}}_{Ih}}|=19932.7 $ nT. In this case the horizontal component of the vector corresponds to the direction North-South with a weak deviation (about 7$^{\circ}$) to the East. Thus, the inclination angle to the magnetic induction vector in the phase center of the radio telescope UTR-2   $ ~\alpha \approx $ 23$^{\circ}$. In a first approximation, the magnetic field of the Earth has a dipole character. Therefore, the value of its magnetic induction decreases with height $ |{\vec{B}_I}| \sim  1/R^3  $, where $R$ is the distance from the center of the magnetic dipole (approximately from the Earth center). Daily variations of Earth's magnetic field for the unperturbed conditions in the middle latitudes do not exceed $ |{\delta \vec{B}_I}| \leqslant 50 \cdot 10^{-9}  $ T. Under these conditions, the parameter $v(F_L) = 0.0574$, $v(F_H) = 0.002617$ ($F_L =$ 23.7 MHz, $F_H =$ 111 MHz).

Substituting the above parameters in equation (\ref{eq4}) we obtain the ellipticity values of the radio emission normal modes in the radio telescope UTR-2 phase center $ M_{O}= -1/{M_{X}} $ at an arbitrary angle between $ \vec k $ and $ \vec B_I $ (see Fig. 6).

From the figure it is clear that both of the ellipticities in the ionosphere in absolute value are not below 0.95 for the considered conditions in the range of angles $ 0^\circ \leqslant {\angle \vec{k} \vec{B}_I} \leqslant  67^\circ ~  \bigcup  ~ 113^\circ  \leqslant  {\angle \vec{k} \vec{B}_I}  \leqslant  180^\circ $ at frequencies $F_c \geqslant$ 23.7 MHz. These ellipticities of normal modes correspond to the case of quasi-longitudinal propagation. Therefore, all further consideration will be carried out from the position of quasi-longitudinal propagation with the circular normal modes.

The same figure shows the ellipticity for normal waves in the pulsar magnetosphere for the value of magnetic induction equal to $ |\vec B_{PSR}|=1000$ G. These values are derived from equations (4). From the calculations it follows that the normal modes will have the quasi-linear polarization in the pulsar magnetosphere at frequency 23.7 MHz for a given value of the magnetic induction in the range of angles $ 20^\circ \leqslant {\angle \vec{k} {\vec{B}}_{PSR}} \leqslant 160^\circ   $ . For the weaker magnetic field in the magnetosphere the normal waves will have elliptical polarization in the whole range of angles, excluding angles close to $0$ and $\pi/2$.

These estimates show that above the frequency of 20 MHz all the layers of the propagation medium (ISM, IPM, EI), except for the interior regions of the magnetosphere of the pulsar, satisfy the conditions of quasi-longitudinal propagation of radio waves.

For quasi-longitudinal electromagnetic $O$ and $X$ waves propagating along the $z$ axis, the eikonal equation can be written as follows:
\begin{equation}                                                          % Eq 5
\begin{array}{l}
\dfrac {d \varphi_{O,X}(\omega)}{dz} =  n_{O,X}(\omega) k(\omega)  =  n_{O,X} (\omega) \dfrac{ {\omega}} {c}
\end{array},
\label{eq5}
\end{equation}

We have used the expansion of the refractive indices in the Taylor series. The members were grouped in such a way that only the first three terms left. They characterize the effects of the propagation medium. Thus the expression that determines the phase of the signal $ \varphi_{O,X}(\omega) $ in any space-time point located on the line of sight was obtained:
\begin{eqnarray}                                                          % Eq 6
%\begin{array}{l}
\varphi_{O,X}(\omega,\psi)  \approx  \omega \dfrac{L} {c}  -
\dfrac{1} {\omega} \dfrac {2 \pi e^{2}} {m_{e} c }
\int \limits _{0} ^{L}  {N_e (z,\psi) }  dz
\nonumber \\
\mp \dfrac {1} {\omega^2} \dfrac{2\pi e^3}  {{m_e}^2 c^2}
\int \limits_{0}^{L}  {N_e (z,\psi) }   \vec B(z,\psi)  d \vec z
%\end{array},
\label{eq6}
\end{eqnarray}
where $L$ is the distance from the pulsar to the observer, $N_e (z,\psi)$ is the electron concentration on the line of sight depending on the pulse phase $\psi$,  $\vec B(z,\psi) d \vec z
= |{ \vec B(z,\psi)}|  \cos({\angle \vec{k} \vec {B}(z,\psi)})d
|{\vec z}|$  is the projection of the magnetic field along the propagation direction depending on the pulse phase.

Equations (\ref{eq7}) and (\ref{eq8}) show separately a phase change caused by the dispersion delay and the Faraday effect, respectively.

\begin{equation}                                                          % Eq 7
\varphi^{D} (\omega,\psi)  \approx  \! \frac{1} {\omega} \frac {2 \pi e^{2}} {m_{e} c }
\int \limits _{0} ^{L} \! {N_e (z,\psi) }  dz = \frac{e^2}{m_e c} DM(\psi) \frac{1} {f}
,
\label{eq7}
\end{equation}
\begin{eqnarray}                                                          % Eq 8
\label{eq8}
{\varphi^R} _{O,X}(\omega,\psi)  \approx \mp  \frac{1} {\omega^2}\frac{2\pi e^3} {m^2 c^2}
\int \limits_{0}^{L} {N_e (z,\psi) }   \vec B(z,\psi)  d \vec z
\nonumber \\
=\mp  RM(\psi) \lambda^2
,
\end{eqnarray}
where $DM(\psi) = \int _{0}^{L} {N_e (z,\psi) } dz$ is the dispersion measure, $RM(\psi) = \frac {e^3}{2 \pi m^2 c^4} \int _{0}^{L} {N_e (z,\psi) }  \vec B(z,\psi)  d \vec z$ is the rotation measure and $f = \omega/(2\pi)$.

A similar approach was used in \citep{1971ApJ...169..487H, 1975MComP..14...55H, 1977ewcp.book.....Z, 1987psr..book.....R} to investigate the possibility of dispersion delay coherent compensation.
We can now go to the matrix model description of the propagation medium. Matrices that are responsible for the delay, caused by dispersion (matrix denoted as $D$) and the Faraday rotation in a cold isotropic plasma in the presence of weak anisotropy (matrix denoted as $ R$) are represented by equations (\ref{eq9}):
\begin{eqnarray}
D= \left [
\begin{array}{ccc}
e^{-i \varphi^D (DM,\omega)} & 0 & 0 \\ 0 & e^{-i
    \varphi^{D}(DM,\omega)} & 0 \\ 0 & 0 & e^{-i\varphi^{D}(DM,\omega)}
\end{array} \right]~,\nonumber % \\
\end{eqnarray}
\begin{eqnarray}                                                                            %eq9
R= \left[
\begin{array}{ccc}
e^{{i\varphi^ {R}}_O(RM,\omega)} & 0 & 0 \\
0 & e^{i{\varphi^ {R}}_{X}   (RM,\omega)}& 0 \\
0 & 0 & 1
\end{array}\right].
\label{eq9}
\end{eqnarray}

Matrix representation of the propagation effects simplifies the construction of the direct and the solution of the inverse problems (see below). In such a representation there is an opportunity to enter as well as to offset the impact of these two propagation effects independently and/or in parts. Accordingly, the inverse problem is solved by matrix inversion used in the construction of the direct problem.

\begin{figure}[h]                                                                       %fig7
%   \imagei
\centering
    \includegraphics[width=64mm]{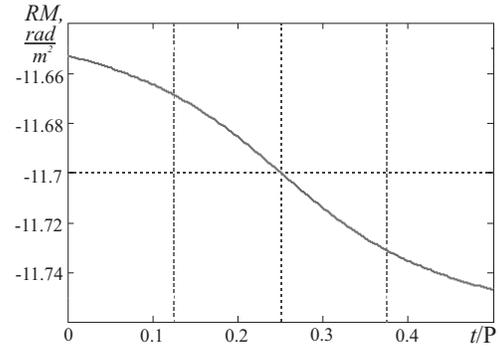}
    \caption{%
        Model change in the value of the rotation measure along the pulse profile.} %% no full stop at the end
    \label{fig7}
\end{figure}

When constructing the direct problem we have abandoned the generally accepted principle, according to which $  RM(\psi)= \text{const} $. In the model representation of the propagation medium, we introduced a rotation measure, which depends on the central frequency and the pulse phase. In the model the value of this parameter for pulse center was chosen equal to the rotation measure of the pulsar B0809+74 (-11.7 rad/m$^2$). To the pulse edges the value varies as the function $ \arctan (m \cdot (\psi-a) ) $ where $m$ is constant (see Fig. \ref{fig7}).

This model dependence of the $RM$ on the radiation longitude was implemented in order to show what kind of restrictions are there in the inverse problem. Rejection of the principle $ RM(\omega_c,\psi)=\text{const} $ illustrates the potential ability for the pulsar magnetosphere resolution in depth.

Below are the equations for the signals that can be detected using WF receivers (registrars of waveform WF signals) \citep{ZakharenkoNikolaenko2007} in a homogeneous layer of the interstellar medium

\begin{eqnarray}                                                      %(10)
\left[
\begin{array}{c}
{{\dot E}_{x}}^{c} (\omega,\psi) \\
{{\dot E}_{y}}^{c} (\omega,\psi) \\
{{\dot E}_{z}}^{c} (\omega,\psi)
\end{array}
\right]= ~~~~~~~~~~~~~~~~~~~~~~~~~~~~~~~~~~~~~~~~\nonumber \\
 \left |
\begin{array}{c}
~ \\ ~ \\ c
\end{array}
\right.
D ~\times~ \left |
\begin{array}{c}
~ \\ ~ \\ b
\end{array}
\right. R^* \times \left |
\begin{array}{c}
~ \\ ~ \\ a
\end{array}
\right.
\left [
\begin{array}{c}
{{\dot E}_{x}}^{a} (\omega,\psi) \\
{{\dot E}_{y}}^{a} (\omega,\psi) \\
{{\dot E}_{z}}^{a} (\omega,\psi)
\end{array}
\right],
\label{eq10}
\end{eqnarray}
where $ {{\dot E}_{x,y,z}}^{c} (\omega,\psi) $  are the incident wave signals in the space below the ionosphere in the phase center of a radio telescope, $ {{\dot E}_{x,y,z}}^{a} (\omega,\psi) $  are the model signals from the equation (\ref{eq3}), $R^*=\dfrac{1} {\sqrt{2} } \left [
\begin{array}{ccc}
1 & 1 & 0 \\ -i & i & 0 \\ 0 & 0 & \sqrt{2}
\end{array} \right] \times R
\times \dfrac {1}{\sqrt{2}}  \left [
\begin{array}{ccc}
1 & i & 0 \\ 1 & -i & 0 \\ 0 & 0 & \sqrt{2}
\end{array}  \right]$ is a sequence of matrices that transforms a linear basis in a circular, then describes the Faraday effect and provides the reverse transition from circular to linear basis. Such a transformation is natural for a weakly anisotropic medium, as the polarization plane rotation is most convenient to be considered in a circular basis (matrix $R$), designations $  \left |
\begin{array}{c}
~ \\ ~ \\ a...c
\end{array}
\right.  $  are the different cross sections (from $a$ to $c$) of the equation (\ref{eq10}).

Equation (\ref{eq10}) shows how the electric field vector is interconnected at the boundaries of a homogeneous layer. As was shown in the papers \cite{1971ApJ...169..487H, 1975MComP..14...55H}, the most simple is to compensate for the impact of the dispersion delay. To compensate this effect coherently it is enough to multiply a matrix, that includes the dispersion delay multipliers, on its inverse matrix. Since the matrix $ D $ characterizing this effect in the chosen frame is diagonal with unit determinant, then the inverse matrix will be a complex conjugate matrix. Note that this approach allows to remove the frequency dispersion delay by parts (by layers) not at once, but gradually to compensate for the phase shift. This may be important at low frequencies. In this area, the dispersion delay can reach minutes and the receiving bands can reach tens of MHz. Therefore, there is no possibility to keep the whole transformed array of numbers in a computer memory.

Similarly, the effect of the polarization plane rotation (i.e., the Faraday effect) is compensated by using multiplication on the inverse to $ R^* $ matrix. In this case the determinants of direct and inverse matrices are equal to one, and direct and the inverse matrices in Cartesian coordinates are Hermitian conjugates. Obviously, the matrices $D$ and $ R^* $ commute. Therefore, in a weakly anisotropic medium compensation of the dispersion delay and the Faraday rotation can be performed in any order.

Let us show how to take into account the influence of the underlying surface on the redistribution of polarization components of the received signal. We assume that the phase centers of linearly polarized dipoles are located on the identical height h in the plane parallel to the underlying surface. Then equation (\ref{eq10}) can be expanded to take into account the interference of the incident and the reflected waves in the receiving direction.

Using the initial arbitrariness in the definition of the position angle in the equations (\ref{eq10}), we define it as follows. In the radio telescope phase center the component $ {{\dot E}_{x}}^{c}(\omega,\psi) $ must be in the incidence plane and conform the vertically polarized component of the incident wave; the component $ {{\dot E}_{y}}^{c} (\omega,\psi) $ must be perpendicular to this plane and parallel to the underlying surface plane (horizontally polarized component of the incident wave); the component $ {{\dot E}_{z}}^{c} (\omega,\psi) $ is directed along the wave vector direction of the incident wave. In a transverse electromagnetic wave , but we will keep this component in the equations for generality. We define the center of LRF in the radio telescope phase center so that the $ x $-axis, $ y $-axis and $ z $-axis coincide with the direction to the South, to the West and into the zenith, respectively.

Vertical and horizontal components of the incident field have different reflection coefficients from the underlying surface. Whereas vertical reflection coefficient must be used with different signs when the vertical component projection of the reflected electric field on the horizontal and vertical vibrators is taken into account. Therefore it is reasonable to introduce three reflection coefficients: $ {\dot {\rho}}_{ps}(\dot \varepsilon_s, \sigma  ,\omega) $, $ {\dot {\rho}}_{p}(\dot \varepsilon_s, \sigma  ,\omega) $ and $ {\dot {\rho}}_{s}(\dot \varepsilon_s, \sigma  ,\omega) $. These reflection coefficients are represented by the following equations:
\begin{eqnarray}                                                                            %eq11
{\dot{\rho}}_{ps}(\dot \varepsilon_s, \sigma  ,\omega)= - \dfrac{\dot \varepsilon _s \sin(\alpha )- \sqrt{\dot \varepsilon_s-{\cos^2 (\alpha )}}}{\dot \varepsilon _s \sin(\alpha )+\sqrt{\dot \varepsilon_s-{\cos^2 (\alpha )}}}
\nonumber \\
{\dot {\rho}}_{s}(\dot \varepsilon_s, \sigma ,\omega) = \dfrac {\sin(\alpha )- \sqrt{\dot \varepsilon_s-{{\cos}^2(\alpha )}}}{ \sin(\alpha )+\sqrt{\dot \varepsilon_s-{\cos^2(\alpha )}}}
~~~\\ \nonumber
{\dot \rho}_{p}(\varepsilon_s, \sigma  ,\omega)= \dfrac{\dot \varepsilon _s \sin(\alpha )- \sqrt{\dot \varepsilon_s-{\cos^2 (\alpha )}}}{\dot \varepsilon _s \sin(\alpha )+\sqrt{\dot \varepsilon_s-{\cos^2 (\alpha )}}}~~~
\label{eq11}
\end{eqnarray}
where $ \dot \varepsilon_s = \Re{[\dot \varepsilon_s]}- i 60\sigma \lambda(\omega ) $ is the complex dielectric permittivity coefficient, $ \Re{[\dot \varepsilon_s]} $ is the real part of it, $ \lambda(\omega ) = 2\pi c/\omega  $ is the wavelength, $ \alpha $ is the elevation angle, $ \sigma $ is the underlying surface conductivity \citep{Aisenberg1985, Sodin1997}, $ {\dot {\rho}}_{ps}(\dot \varepsilon_s, \sigma  ,\omega) = -{\dot {\rho}}_{p}(\dot \varepsilon_s, \sigma  ,\omega) $ .

Except for the difference in the reflection coefficients, it should be noted that the incident and the reflected waves are summarized with different phases. It is caused by the fact that each of them will pass the different path to the receiving dipole phase center. Thus, the influence of the underlying surface can be modeled by introducing a matrix that takes into account the interference of incident and reflected waves in the plane of incidence and the plane perpendicular to it:

\begin{gather}                                                              %eq12
\left[
\begin{array}{ccc}
1& 0 & 0 \\
0 & 1& 0 \\
0 & 0 & 1
\end{array} \right] + \nonumber \\
 \left[
\begin{array}{ccc}
-{\dot \rho}_{p} (\dot \varepsilon, \sigma, \omega) e^{-i \phi} & 0 & 0 \\
0 & {\dot \rho}_s (\dot \varepsilon, \sigma, \omega) e^{-i \phi}& 0 \\
0 & 0 & {\dot \rho}_{p} (\dot \varepsilon, \sigma, \omega) e^{-i \phi}
\end{array} \right] \nonumber \\
~~~~~~~~~~~~~~~~~~~~~~~~~~~~~~~~~~~~~~~~~~~~~~~~~=UN + REF ~,
\label{eq12}
\end{gather}
where $ \phi = \omega \frac{2 h}{c} \sin(\alpha) $ is the phase shift between the incident and reflected signal, $h$ is the height of the dipole phase center above the underlying surface.
It should be also considered the redistribution of the incident radiation component in the observer reference frame at the registration. To do this, we write the matrix of the direction cosines tied to the radio telescope UTR-2:
\begin{gather}                                                      %eq13
COS_1 \cdot COS_2= \nonumber\\
\left[\begin{array}{ccc} \cos(\theta) &-\sin(\theta) & 0 \\ \sin(\theta) & \cos(\theta) &0 \\
0 & 0 & 1 \end{array}\right] \cdot %\nonumber \\
\left[\! \! \begin{array}{ccc} -\sin(\alpha) & 0 & \cos(\alpha) \\ 0 & -1 & 0\\
-\cos(\alpha) & 0 & -\sin(\alpha) \end{array} \!\!\right] \nonumber \\=
\left[\begin{array}{ccc} -\sin(\alpha)\cos(\theta) &\sin(\theta) & \cos(\alpha)\cos(\theta) \\ -\sin(\alpha)\sin(\theta) & -\cos(\theta) & \cos(\alpha)\sin(\theta) \\
-\cos(\alpha) & 0 & -\sin(\alpha) \end{array}\right],
\label{eq13}
\end{gather}
where $\theta$ is the azimuth angle (the positive values of $ \theta $ are measured from the direction of the South, through the West to the North) \citep{Kulikov1976, igrf11_2013}.
In accordance with the International Astronomical Union recommendations, the coordinate system, connected with the telescope, is a left-handed triplet of vectors ($COS_1$). The source coordinate system, chosen by us, is also the left triple ($COS_2$).

Assume that the radio telescope contains three groups of the thin linear dipoles, the polarization of which corresponds to the selected basis vectors of LRF and dipoles themselves do not interact with each other. The diagonal normalization matrix of the received signal by the effective area of the radio telescope \citep{1970pewp.book.....G,1987tfia.book.....G,1997riap.book.....Z} has the form:
\begin{eqnarray}                                                                    %eq14
EFS=
\left[
\begin{array}{ccc}
N_x^{\text{\textit{eff}}}  & 0 & 0 \\ 0 & N_y^{\text{\textit{eff}}} & 0 \\ 0 & 0 & N_z^{\text{\textit{eff}}}
\end{array} \right] = \nonumber ~~~~~~~~~~~~~~~\\
~~~~~~~~~~~~~~~~~~~~\left[
\begin{array}{ccc}
S_x^{\text{\textit{eff}}} / S_x^{d}  & 0 & 0 \\ 0 & S_y^{\text{\textit{eff}}} / S_y^{d} & 0 \\ 0 & 0 & S_z^{\text{\textit{eff}}} / S_z^{d}
\end{array}
\right].
\label{eq14}
\end{eqnarray}
where the diagonal element $S_{x,y,z}^{\text{\textit{eff}}}$ is the ratio of the radio telescope effective area  to the single dipole effective area $S_{x,y,z}^{d}$ in the selected linear polarization ($x; y; z$). For example, $ EFS_{(x,x)}=S_{x}^{\text{\textit{eff}}} / S_{x}^{d} = N_{x}^{\text{\textit{eff}}} $  where $ N_{x}^{\text{\textit{eff}}} $ is the effective number of dipoles in a specific radio telescope with the main axis aligned along the x axis. It is assumed that the effective area of the radio telescope and dipoles for each polarization are known in the operating frequency band in either pointing direction.

Based on the expressions above, equation (\ref{eq10}) is transformed into equation (\ref{eq15}):
\begin{eqnarray}
\left[                          %eq{15}
\begin{array} {c}
{\dot E}_{x}^g (\omega,\psi ) \\ {\dot E}_{y}^g (\omega,\psi )
\\ {\dot E}_{z}^g (\omega,\psi )
\end{array} \right] =
\left |
\begin{array}{c}
~ \\ ~ \\ g
\end{array}
\right.  \!\!\! S_{eff} \times
\left |
\begin{array}{c}
~ \\ ~ \\ f
\end{array}
\right. \!\!\!COS_1 \times
\left |
\begin{array}{c}
~ \\ ~ \\ e
\end{array}
\right. \!\!\! COS_2 \nonumber \\ \times
\left |
\begin{array}{c}
~ \\ ~ \\ d
\end{array}
\right. (UN + REF) \times
\left |
\begin{array}{c}
~ \\ ~ \\ c
\end{array}
\right.  \left [
\begin{array}{c}
{\dot E}_x^c (\omega,\psi ) \\ {\dot E}_y^c (\omega,\psi ) \\
{\dot E}_z^c (\omega,\psi )
\end{array} \right ].
\label{eq15}
\end{eqnarray}
where $ {\dot E}_{x,y,z}^g (\omega,\psi ) $ is the complex amplitude at the "average" linear vibrator output in the laboratory reference frame (LRF or further the cross-section $"g"$). LRF axes ($x, y, z$) are directed from the phase centre of RT to the South, to the West, and to the zenith, respectively. LRF center coincides with the radio telescope phase center.

Equation (\ref{eq15}) represents a model signal passed trough the propagation medium and received by the considered radio telescope. The presented model of the polarized signal and the propagation medium can be used to develop methods of determining the polarization parameters.\textbf{}

\section{Methods for determination the polarization parameters and the propagation medium parameters.}

One can trace step by step the changes that occur with the polarization parameters of the signal as it passes through the medium.

For determining the Stokes parameters, we use equations (\ref{eq16}) and (\ref{eq17}). These equations associate the electric field vectors with the polarization tensor J and the Stokes parameters $I, Q, U, V$ in a circular and a linear basises:
\begin{eqnarray}                                                                        %eq16
J(\omega, \psi ) = \left [
\begin{array}{cc}
I_{xx}(\omega, \psi) & I_{xy}(\omega, \psi) \\ I_{yx}(\omega, \psi) & I_{yy}(\omega, \psi)
\end{array}  \right]=
\frac{c} {4\pi} \cdot \nonumber \\
\left ( \!\!\!
\begin{array}{cc}
\langle \dot {E}_x (\omega, \psi) \cdot  \overline {\dot {E}_x (\omega, \psi)} \rangle_t & \langle \dot {E}_x (\omega, \psi) \cdot  \overline {\dot {E}_y (\omega, \psi)} \rangle_t  \\ \langle \dot {E}_y (\omega, \psi) \cdot \overline { \dot {E}_x (\omega, \psi)} \rangle_t  & \langle \dot {E}_y (\omega, \psi) \cdot  \overline { \dot {E}_y (\omega, \psi)} \rangle_t
\end{array} \!\!\! \right )  ,
\label{eq16}
\end{eqnarray}
\begin{eqnarray}                                                                        %eq17
\left\{
\begin{array}{l}
I(\omega, \psi)= I_{xx}(\omega, \psi) +I_{yy}(\omega, \psi)  \\ Q(\omega, \psi) = I_{xx}(\omega, \psi)-I_{yy}(\omega, \psi) \\ U(\omega, \psi) = I_{xy}(\omega, \psi) + I_{yx}(\omega, \psi) \\ V(\omega, \psi)= i(I_{yx}(\omega, \psi) -I_{xy}(\omega, \psi))
\end{array}
\right.
 \nonumber , \\
\left\{
\begin{array}{l}
I(\omega, \psi)= I_{rr}(\omega, \psi) +I_{ll}(\omega, \psi)  \\ Q(\omega, \psi) = I_{lr}(\omega, \psi)-I_{rl}(\omega, \psi) \\ U(\omega, \psi) =  -i (I_{rl}(\omega, \psi) - I_{lr}(\omega, \psi)) \\ V(\omega, \psi)= I_{rr}(\omega, \psi) -I_{ll}(\omega, \psi)
\end{array} \right.
\label{eq17}.
\end{eqnarray}
Here under the polarization tensor element, for example $ \langle \dot {E}_x (\omega, \psi) \cdot  \overline {\dot {E}_x (\omega, \psi)} \rangle_t $, we understand the time-averaged value of the complex amplitude $ \dot {E}_x (\omega, \psi) $ and complex conjugate amplitude $ \overline {\dot {E}_x (\omega, \psi)} $ (the operator $  \langle * \rangle_t  $ defines the averaging). Variation of the amplitude with time is assumed to be slow, such that its value remains constant at a characteristic time interval $\Delta\tau \gg 1/\Delta\omega$ (here $ \Delta\omega $ is the recorded frequency band).
The following polarization parameters are the position angle $ \chi(\omega,\psi) $, the angle $\xi(\omega,\psi)$, which characterizes the ratio of the semi-major axes of the polarization ellipse, and the ellipticity $ \varepsilon(\omega,\psi) $. These parameters can also be determined from the Stokes parameters:
\begin{gather}                                                                              %eq18
\chi(\omega, \psi) = \frac{1} {2} \arg ({Q(\omega, \psi) + iU(\omega, \psi) })\notag,
\\
\xi(\omega, \psi) =\frac{1} {2} \arcsin \!\! \left[\!{\frac{V(\omega, \psi)}{\sqrt{ ({Q^2(\omega, \psi) \!+\! U^2(\omega, \psi) \!+\! V^2(\omega, \psi) })}}}\!\right]\notag \!\!,
\\
\varepsilon(\omega, \psi)  =\tan(\xi(\omega, \psi) ).
\label{eq18}
\end{gather}
Relative polarization parameters such as $P_t$ is the total polarization degree, $P_l$ is the degree of linear polarization, $P_c$ is the degree of circular polarization, $P_d$ is the depolarization degree, are defined as follows:
\begin{gather}                                                                          %eq19
P_{t}(\omega, \psi) = \frac{\sqrt{Q^2(\omega,\psi)+U^2(\omega,\psi)+V^2(\omega,\psi)}} {I(\omega,\psi)} \notag,\\
P_{l}(\omega,\psi) = \frac{\sqrt{Q^2(\omega,\psi)+U^2(\omega,\psi)}} {I(\omega,\psi)}\notag,\\
P_{c}(\omega,\psi) = \frac {V(\omega,\psi)} {I(\omega,\psi)}\notag,\\
\sqrt{{P_t}^2(\omega,\psi) + {P_l}^2(\omega,\psi) + {P_c}^2(\omega,\psi) + {P_d}^2(\omega,\psi)} = 1.
\label{eq19}
\end{gather}
It is notable that when assessing polarization parameters of the pulsed signals, one cannot accumulate the Stokes parameters for a long time (similar to how it is done in the optical range). In case of registration of pulsed signals the sensitivity will be limited by the time resolution required for the task solution.

One of the acceptable options for increasing the sensitivity of polarization measurements is associated with accumulation of the Stokes parameters in the frequency range. You can use the fact that the Stokes parameters in the reference frame of radiation source are identical in a relatively narrow frequency range (see Fig. \ref{fig2} and Eq. (\ref{eq1}, \ref{eq2}, \ref{eq10})). However, a similar accumulation of the Stokes parameters in frequency in LRF leads to distortions in the initial polarization parameters estimates.

\begin{figure}[h]                                                               %fig8
    %\imagei
    \centering
    \includegraphics[width=70mm]{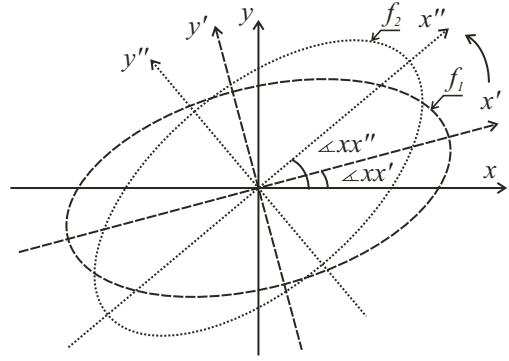}
    \caption{%
    Faraday rotation of the polarization plane in LRF for different radiation frequencies ($f_1, f_2$). In the figure the coordinate axes ($x, y$) correspond to the major axes of the polarization ellipse in a narrow frequency band in the reference frame associated with the radiation source. Similar coordinate axes ($x', y'$) and ($x'', y''$) correspond to the principal axes of the polarization ellipse at frequencies $f_1$ and $f_2$ in LRF.} %% no full stop at the end
    \label{fig8}
\end{figure}

In particular, not only the parameters $ \langle {Q^g(\omega, \psi)} \rangle_\omega  $ and $ \langle {U^g(\omega, \psi)} \rangle_\omega  $  are distorted, but the parameter $ \langle {{\sqrt{(Q^g)^2(\omega,\psi)+(U^g)^2(\omega,\psi)}} } \rangle_\omega  $ is not the invariant. This is due to the Faraday effect, i.e. rotation of the polarization plane on the propagation path (Fig. \ref{fig8}).

Therefore, in the case of the pulsed signals registration the systems with memory should be used. These systems, on the one hand, can reduce the local reception band $ \Delta\omega $ to a width where the depolarization of a linearly polarized component caused by the Faraday effect is low, i.e. the following relations are satisfied:
\begin{gather}                                                                      %eq20
\int \limits_{\omega_{c}-\Delta \omega/2} ^ {\omega_{c}+\Delta \omega/2} {\sqrt{(Q^c(\omega,\psi))^2+(U^c(\omega,\psi))^2}}  d \omega  \approx  \nonumber \\
~~~~~~~~~~~~ \int \limits_{\omega_{c}-\Delta \omega/2} ^{\omega_{c}+\Delta \omega/2} {\sqrt{(Q^a(\omega,\psi))^2+(U^a(\omega,\psi))^2}}  d \omega ,
\label{eq20}
\end{gather}
where $ \Delta\omega $ is the registrator spectral resolution, $ \omega_c $ is the center frequency for the local reception band, $  Q^c( \omega, \psi) ,~ U^c (\omega, \psi)  $ are the corresponding Stokes parameters in the "free" space at a distance equal to the distance from the radiation source to the receiver (excluding the effect of the underlying surface, see equation (\ref{eq10})), $  Q^a( \omega, \psi) ,~ U^a (\omega, \psi)  $ are the similar Stokes parameters of the source reference frame.

Thus the narrow band response to the linearly polarized radiation of a parabolic steerable radio telescope that is not normalized to the effective area (see eq. (\ref{eq15})) is given by equation (\ref{eq20}). In the case of meter and decameter radio telescopes, the equation (20) represents the same response only for a hypothetical radio telescope, the dipole plane of which is at the same distance as its phase center, perpendicular to the wave vector $ \vec k (\omega_c) $. In this case the underlying surface does not affect the response of such a radio telescope. Figure 8 shows that the rotation of the plane of polarization between the extreme frequencies in the band $ \Delta\omega $ should be about 1○  one degree taking into account the conditions (\ref{eq20}).

Assume that in cross section $"c"$ of the equation (\ref{eq15}) we have been able to compensate the dispersion delay influence and moved to section $"b"$. In this case, the Stokes parameters $I^b(\omega,\psi)$, $Q^b(\omega,\psi)$, $U^b(\omega,\psi)$, $V^b(\omega,\psi)$, received in the narrow bands in the "free" space, the following relations will be fair :
\begin{gather}                                                              %eq21
 \left[ \begin{array}{c}
  I^{b}(\omega,\psi) \\ Q^{b}(\omega,\psi) \\ U^{b}(\omega,\psi) \\ {V}^{b}(\omega,\psi) \end{array}
  \right]   = \nonumber \\
    \left[ \begin{array}{cccc}
    1 & 0 & 0 & 0 \\ 0 & ~~\cos(2\varphi_{RM}(\omega,\psi)) & \pm \sin(2\varphi_{RM}(\omega,\psi)) & 0 \\ 0 & \mp \sin(2\varphi_{RM}(\omega,\psi))  &~~ \cos(2\varphi_{RM}(\omega,\psi)) & 0 \\ 0 & 0 & 0 & 1
    \end{array}  \right] \cdot\nonumber \\
    ~~~~~~~~~~~~~~~~~~~~~~~~~~~~~~~~~~~~~~~~~~~~\left[
    \begin{array}{c} I^a(\omega,\psi) \\ Q^a(\omega,\psi) \\ U^a(\omega,\psi) \\ V^a(\omega,\psi) \end{array} \right], \nonumber \\
    \left[ \begin{array}{c}
        I^{a}(\omega,\psi) \\ Q^{a}(\omega,\psi) \\ U^{a}(\omega,\psi) \\ {V}^{a}(\omega,\psi) \end{array}
    \right]   = \nonumber \\
    \left[ \begin{array}{cccc}
        1 & 0 & 0 & 0 \\ 0 & \cos(2\chi(\omega_c,\psi)) & - \sin(2\chi(\omega_c,\psi)) & 0 \\ 0 &  \sin(2\chi(\omega_c,\psi))  & ~~\cos(2\chi(\omega_c,\psi)) & 0 \\ 0 & 0 & 0 & 1
    \end{array}  \right] \cdot\nonumber \\
    ~~~~~~~~~~~~~~~~~~~~~~~~~~~~~~~~~~~~~~~~~~~~\left[
    \begin{array}{c} I^0(\omega,\psi) \\ Q^0(\omega,\psi) \\ 0 \\ V^0(\omega,\psi) \end{array} \right],
    \label{eq21}
\end{gather}
where $ \varphi_{RM}(\omega,\psi)  =  RM(\psi,\omega) \cdot (2\pi c/\omega )^2 $ is the polarization plane rotation angle, $ {I}^{0}(\omega,\psi)$ ,  $Q^{0}(\omega,\psi)$ , ${V}^{0}(\omega,\psi) $ are the Stokes parameters in the source reference frame two axes of which coincide with the polarization ellipse major axes,  $ {I}^{a}(\omega,\psi)$,  $Q^{a}(\omega,\psi)$, $U^{a}(\omega,\psi)$, ${V}^{a}(\omega,\psi) $ are the same parameters in the source reference frame the axes of which are rotated with respect to the polarization ellipse major axes by an arbitrary position angle $ \chi (\omega_c, \psi ) $.

It follows from the discussion above that averaging of the Stokes parameters in a narrow band of frequencies can be carried out for the signal only after bringing it into source RF. To do this using the meter and decameter radio telescopes, two steps will be needed to perform. First, the signal, registered in a narrow band, should be brought in RF associated with the "free" space. Thus we get rid of the distortions introduced by the underlying surface in the received signal. This step is quite complicated. It is associated with the type of recorder at the receiving side. Therefore, this step will be discussed in the Part II. Secondly, the polarization parameters in RF associated with "free" space should be brought in a RF associated with the source, in accordance with equation (\ref{eq21}). This requires to reverse the equation (\ref{eq21}). However, for this we should know the value of $  RM(\omega_c, \psi) $. Initially, it is assumed that the value  $  RM(\omega_c, \psi) $ is unknown to us.

\subsection[Determination of the rotation measure for parabolic radio telescopes.]{Determination of the rotation measure for parabolic radio telescopes.}

Let us consider how to determine $  RM(\omega_c, \psi) $. A registered response $  \langle \dot {E}_x  (\omega,\psi) \cdot  \overline {\dot {E}_x  (\omega,\psi)} \rangle_t  =   | {\dot {E}_x  (\omega,\psi)}|^2 $ is considered to be the original, whose arguments are the angular frequency $\omega$  (or the regular frequency $f$) and pulse phase $ \psi $ (see Fig. \ref{eq9}a,b). We believe the signal in system $"b"$ is the original, because the methods for determining and compensation the frequency dispersion delays are well known and have a sufficiently high accuracy. Methods of the underlying surface influence compensation and the accounting of redistribution polarization components of electric field will be considered in the future paper. These methods will be important for radio telescopes such as the UTR-2, URAN-2, LOFAR, BSA, GEETEE, which are located above the surface of the Earth or above a metal grid.

There are several methods for determining the rotation measure. In one case \citep{2013BaltA..22...53U}, the response is expanded in a Fourier series. In another case by method minimum mean square deviations (MMSD) of the sinusoid, whose variable parameters are amplitude, frequency and initial phase, is being fit into the given response \citep{1974AZh....51..927S,1983SvA....27..322S,1988SvA....32..177S,1989puls.book...42S}. The first method is preferable because it automatically allows us to estimate the amplitude, frequency and phase of the polarization response to the linearly polarized component of the signal. In both methods, the value $  RM(f_c, \psi) $ is estimated from the equation:
\begin{align}                                                               %eq22
|{RM_{est}(f_c, \psi)} | =  \frac{\pi}{c^2} \cdot \left [\frac{f_c^2 (f_c +
    \Delta f_F(\omega_c ,\psi ))^2 }  {( f_c + \Delta f_F(\psi )
    )^2 -f_c^2 }\right ] \nonumber \\
 \approx \frac{ \pi f_c^3} {2 c^2 \Delta f_F(\psi )},
 \label{eq22}
\end{align}
where $ |{RM_{est}(f_c, \psi)} | $ is the RM estimation at the pulse phase $ \psi $, $ f_c $ is the central receiving frequency, $ \Delta f_F(\psi) $ is the Faraday modulation period in a given frequency band, corresponding to the presence of a linearly polarized component in the radiation (see Fig. \ref{eq9}$ c $).
\begin{figure}[h]                                                               %fig9
%   \imagei
    \includegraphics[width=84mm]{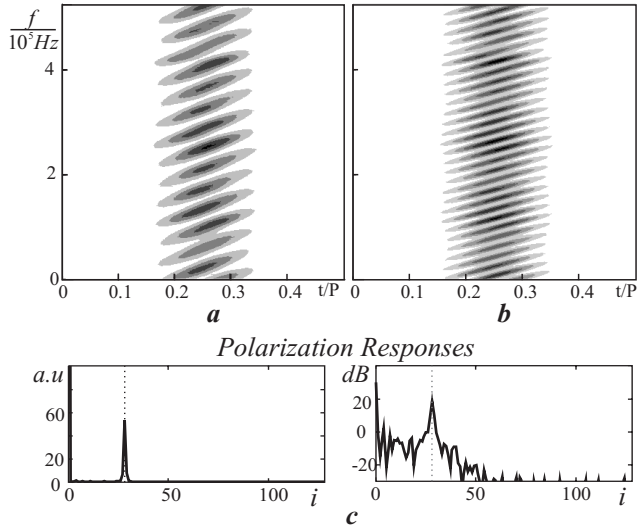}
    \caption{%
    Rotation Measure determination. a) and b) are the model signal dynamic spectra at central receiving frequencies $ f_c= 30$ MHz and $ f_c= 23.7$  MHz respectively; c) is the spectral decomposition of the model signals $ \dot P(t_i,a )  =  FFT \left[ {| {\dot {E}_x  (\omega,a)}}|^2 \right] $ at $ f_c=23.7 $ MHz and central pulse longitude $ \psi=a $ in the linear power scale (arbitrary units) and in decibels.} %% no full stop at the end
    \label{fig9}
\end{figure}

In \cite{2013BaltA..22...53U} the position angle estimation $ \chi_{est}(\omega_c,\psi) $ was also obtained from the argument of the function  $ \dot P(t_i,a_j )  =  FFT \left[ {| {\dot {E}_x  (\omega_i,a_j)}}|^2 \right] $:
\begin{align}                                                           %eq23
~~~~~~~~~~~~~\chi_{est}(\omega_c,\psi)~= \frac {1}  {2} \arg ({\dot P(t_{max} ,\psi ) }),
\label{eq23}
\end{align}
where $ t_{max} $ is the Fourier transform frequency of the analyzed signal (with the dimension of time), at which the value $ P(t ,\psi )  $ is maximum (see Fig. \ref{eq9}c).

From the above reasoning is clear that $ \Delta f_F( \omega_c,\psi ) \sim 1/({t_{max}(\psi)})  $. Accordingly, methodological errors of this approach $ \delta RM(\omega_c,\psi)/RM(\omega_c,\psi)  =  1/t_{max}(\psi) $ will be inversely proportional to the dimensionless frequency $ t_{max}(\psi ) = n_{max}(\psi ) $. Considering the equations (\ref{eq10}, \ref{eq15}) it is clear that the use of a single sine/cosine functions as a model representation of the polarization response at the receiving side does not adequately represent the transformation of polarized signals in the propagation medium.

In this paper we propose to take the next step, which will enhance the achieved methodological precision in determination of $ RM_{est}(\omega_c,\psi ) $ even by $ \sqrt {Nq} $ times, where $ Nq $ is a number of Nyquist frequency in the discrete Fourier transform of the analysed signal $  {| {\dot {E}_x  (\omega,\psi)}}|^2 $. This improvement allows to bring the relative accuracy in determining $ RM_{est}(\omega_c,\psi ) $ up to the current accuracy in determining of $ DM $ at decameter waves. This, in turn, gives the potential opportunity for probing deep into upper layers of the pulsar magnetosphere.

Algorithm capable of improving the accuracy of $ RM $ estimates is based on the above models of the polarized pulsed radiation and its propagation medium model. Thus, evaluation of $ {\dot E}_x (\omega,\psi )  $ and $ {\dot E}_y (\omega,\psi )  $, made on the basis of analysis of the signals $ {\dot E}_x^c (\omega,\psi )  $ and $ {\dot E}_y^c (\omega,\psi )  $ or only one signal $ {\dot E}_x^c (\omega,\psi )  $, can be substituted into the right-hand side of equation (\ref{eq10}).

Below we consider the most complicated case when the response is known only in one channel. Obtained in (\ref{eq22}) and (\ref{eq23}) values of $ RM_{est}(\omega_c,\psi ) $ and $ \chi_{est}(\omega_c,\psi) $ can be regarded as the impact parameters for MMSD method. Next you need to minimize the discrepancy $ U(RM(\omega_c,\psi),\chi (\omega_c,\psi)) $ in each pulse phase. And the value $ RM(\omega_c,\psi ) $ should be changed in the range $ RM_{est}(\omega_c,\psi)(1 \mp  2/ t_{max}(\omega_c,\psi)) $ as the parameter with a step $ RM_{est}(\omega_c,\psi)/( t_{max} (\omega_c,\psi) \sqrt{Nq})$, and the value $ \chi(\omega_c,\psi) $ - in the range $[-\pi, \pi]$ with a step $ RM_{est}(\omega_c,\psi) c^2/(f^2 _L t_{max} (\omega_c,\psi) \sqrt{Nq}) $:
\begin{gather}                                %eq.24
U(RM(\omega_c,\psi ),\chi (\omega_c,\psi))= \nonumber \\
~~~~~~~~~~~~~~~~~~\sum _{i=0}^{Nq} (|{{\dot {E}}_x^{c}(\omega_i,\psi )}|
- |{{\dot {E}}_x^{mod}(\omega_i,\psi ) }|)^2~,
\label{eq24}
\end{gather}
where $ U(RM(\omega_c,\psi ),\chi (\omega_c,\psi)) $ is the discrepancy depending on the desired parameters,  $ |{\dot E}_x^c (\omega,\psi )|  $  is actually registered in the polarization channel A intensity in section $"c"$, $ |{{\dot {E}}_x^{mod}(\omega_i,\psi ) }| $ is the intensity of model signal in the same channel, obtained from equation (\ref{eq10}).

\begin{figure}[h]                                                               %fig10
    %\imagei
    \includegraphics[width=84mm]{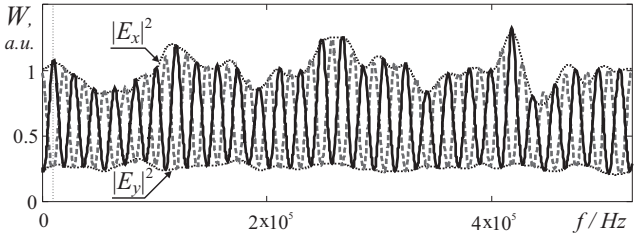}
    \caption{%
        The power spectral density of linearly polarized dipole located in the "free" space. Solid curve is the signal $ {|{{\dot {E}}_x^{c}(\omega,a )}|}^2  $, dashed line is the virtual signal $ {|{{\dot {E}}_y^{c}(\omega,a )}|}^2  $, $ \psi = a  $ is the phase, corresponding to the model pulse center. The power spectral density $ {|{{\dot {E}}_x(\omega,a )}|}^2  $ of the linear polarization A in the radiation source reference frame (dotted line above (envelope)); it is proportional to the square of the module length of the polarization ellipse semi-major axis. Similarly, the power spectral density $ {|{{\dot {E}}_y(\omega,a)}|}^2  $ of the linear polarization B in the radiation source reference frame (dotted line below); it is proportional to the square modulus length of the polarization ellipse semi-minor axis. } %% no full stop at the end
\label{fig10}
\end{figure}

The retrieved parameters $ RM_{est2}(\omega_c,\psi ) $ and  $ \chi_{est2}(\omega_c,\psi) $ will correspond to the minimum of the discrepancy $ \min [{U(RM_{est2}(\omega_c,\psi),\chi_{est2} (\omega_c,\psi))}]  $. It remains to show how to obtain from the signal $ |{{\dot {E}}_x^{b}(\omega,\psi)}|  $ the estimates $ {\dot E}^0_x (\omega,\psi )  $ and $ {\dot E}_y^0 (\omega,\psi )  $ required for the proposed algorithm implementation. Let us take into account that in the "free" space due to the Faraday effect, polarization ellipses for different frequencies have different position angles relative to RF in which they are recorded (see Fig. \ref{fig8}). Energy of initial linearly polarized components in the observer reference frame is redistributed between the relevant polarizations depending on the frequency, as shown in Figure \ref{fig10}. Moreover, the total energy in the two orthogonal polarizations in any cross section of the equation (\ref{eq10}) does not change if the radio telescope plane, where the dipoles are located, is perpendicular to the radiation direction. This is a simple consequence of the energy conservation law. In the equation (\ref{eq10}) it is displayed that the determinants of transforming matrices $ D $ and $ RM^* $ are equal to 1. Hence the envelope drawn by the peak values of the power spectral density registered in a linearly polarized channel from the observer reference frame is proportional to the power spectral density of the polarization ellipse semi-major axis defined in the source reference frame. I.e., $ {|{{\dot {E}}_x^{b}(\omega_{lmax},\psi_i )}|}^2   \approx  {|{{\dot {E}}_x^0(\omega_{lmax},\psi_i )}|}^2   $ where $ \omega_{lmax} $ are the frequencies of local maxima of function. Accordingly, the envelope drawn for the minimal values of the power spectral density, registered in the same conditions is proportional to the power spectral density of the polarization ellipse semi-minor axis, defined in the radiation source reference frame $ {|{{\dot {E}}_x^{b}(\omega_{lmin},\psi_i )}|}^2   \approx  {|{{\dot {E}}_y^{0}(\omega_{lmin},\psi_i )}|}^2   $. Here $ \omega_{lmin} $ are the frequencies of local minima of the function $ {|{{\dot {E}}_x^{b}(\omega,\psi )}|}^2  $ (see Fig. \ref{fig10}).

Since on the reception side one can obtain good estimations of $ {|{{\dot {E}}_{x~ est}^{0} (\omega,\psi )}|}^2  $ and $ {|{\dot {E}}_{y~ est}^{0}(\omega,\psi )|}^2  $ only in the nodes corresponding to the frequencies $ \omega_{lmax};~  \omega_{lmin}   $ the estimates can be performed by any iterative algorithms in the remaining frequency band. Here, the assumption is that the functions $ {|{{\dot {E}}_x^{0}(\omega,\psi )}|}^2  $ and $ {|{{\dot {E}}_y^{0}(\omega,\psi )}|}^2  $ are sufficiently smooth (see Fig. \ref{fig2}).

The number of the nodes $ \omega_{lmax};~  \omega_{lmin}   $ can be increased quite simply, and thereby, improve the required estimates. For this purpose the virtual channel B is created from a single, available on the reception side channel A. At the first step, two Fourier images of the signal, shifted by the complex conjugate phase factors ($ \exp(\pm i \Omega \omega_{sh}(\omega_c,\psi) ) $) are obtained using an FFT of the signals $ {|{{\dot {E}}_x^{b}(\omega,\psi )}|}^2  $ in the channel A:
\begin{gather}                                                          %eq25
\dot {S\!p}_{+sh;\!-sh}(\Omega,\! \psi) = F\!F\!T[ {|{{\dot {E}}_x^{b}(\omega,\!\psi )}|}] \cdot \! \exp(\pm i \Omega \omega_{sh}(\omega_c,\!\psi) ) .
\label{eq25}
\end{gather}
In this equation $ \omega_{sh}(\omega_c,\psi) = \Delta \omega  /(2 \cdot n_{max}(\omega_c,\psi)) $ has the dimension of Hz (frequency shift i.e. shift in the argument of the original), $ n_{max}(\omega_c,\psi) $ is the number of non-zero harmonic with a maximum power spectral density of the signal (see Fig. \ref{fig9}$ c $).

At the second step the original signal $ {| {{\dot E}_{y~ virt}^{b}( \omega, \psi)}|}^2 $ is generated from the obtained image signals in the virtual channels $(+sh;-sh)$ using the inverse FFT and by averaging of virtual polarization channel B:
\begin{gather}                                                          %eq26
{| {{\dot {E}}_{y~ virt}^{b}(\omega,\psi)}|}^2 = \dfrac{I\!F\!F\!T[ {\dot S\!p}_{+sh}( \Omega, \psi) +  {\dot S\!p}_{-sh}(\Omega, \psi) ]} {2}  .
\label{eq26}
\end{gather}
This two-step procedure can be repeated many times by setting the condition $ \tau_{sh_j}(\omega_c,\psi) = \tau_{sh_{j-1}}(\omega_c,\psi)/2  $ on the j-th iteration step. In the present numerical simulations such iteration was held 4 times. Next, using a peak detector we got the upper envelope $ {| {\dot E_{H}( \omega, \psi)}|}^2 $ of signals in the real and virtual channels in the whole band. Similarly, the lower envelope $ {| {\dot E_{L}( \omega, \psi)}|}^2 $ in the same channels, we obtained using an anti-peak detector.

\begin{figure}[h]                                                               %fig11
    %\imagei
    \includegraphics[width=74mm]{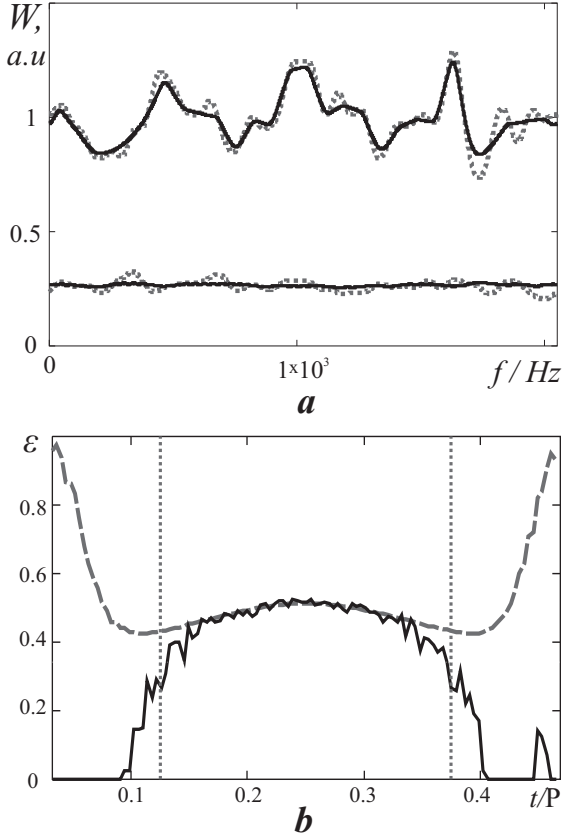}
    \caption{%
        Comparison of the power spectral densities with the original values of the same quantities in the polarization channels A and B in the source reference frame. $ a) $ is the original signals $ {|{{\dot {E}}_x(\omega,a )}|}^2  $ and $ {|{{\dot {E}}_y(\omega,a )}|}^2  $ (upper and lower dotted curves, respectively), where $a$ is the pulse phase center. Solid upper and lower curves are signals $ {| {\dot E_{H}( \omega, a)}|}^2 $ and $ {| {\dot E_{L}( \omega, a)}|}^2 $, respectively. The mentioned signals are our estimates of signals $ {|{{\dot {E}}_x(\omega,a )}|}^2  $ and $ {|{{\dot {E}}_y(\omega,a )}|}^2  $ according to the data recorded on the reception side in the "free" space. $b)$ is comparison of the average ellipticity (solid curve) with model values of this factor (dashed curve) at all pulse phases. } %% no full stop at the end
    \label{fig11}
\end{figure}

The resulting dependencies are shown in Fig. \ref{fig11} and compared with the same initial dependencies, which are known by the numerical simulations results. In the same figure (Fig. \ref{fig11}$ b $) the comparison of the average ellipticity estimates $ \langle \varepsilon_{est} (\omega_c,\psi) \rangle_\omega  $ and its original dependence is given. Estimate of the average ellipticity is obtained from the expression:
\begin{gather}                                                          %eq27
\langle \varepsilon_{est} (\omega_c,\psi) \rangle_\omega  = \dfrac{1}{Nq}  \cdot \sum \limits _{i=0} ^{Nq-1} \sqrt{\dfrac{{{| {\dot E_{L}( \omega_i, \psi)}|}^2} }  {|{\dot E_{H}( \omega_i, \psi)|}^2} } .
\label{eq27}
\end{gather}

It can be seen that the envelope of the signal in channels A and B adequately reflects our intuitive understanding of both the model properties and the signals. I.e. our estimates are the better the higher is the ratio $S/N$.
Now the restoration of the original signals in the source reference frame does not cause difficulties:
\begin{gather}                                                                  %eq28
    {\dot {E}}^0_{x~ est}(\omega,\psi ) = \sqrt{|{{\dot {E}}_{H}(\omega,\psi )|}^2} , \nonumber \\
        {{\dot {E}}^0_{y~ est}(\omega,\psi )} = \langle \varepsilon_{est} (\omega_c,\psi) \rangle_\omega  \cdot {{\dot {E}}_{x~ est}(\omega,\psi )} \cdot \exp(-i \dfrac{\pi}{2} ) ,
        \label{eq28}
\end{gather}
where ${\dot {E}}^0_{x~ est}(\omega,\psi )$ and ${\dot {E}}^0_{y~ est}(\omega,\psi )$ are the restored complex amplitudes of the polarization ellipse in the source reference frame.

Clearly, we were able to restore the complex amplitudes of up to an unknown phase factor $  \exp(-i \varphi(\omega )) $ that is identical for both amplitudes, where $ \varphi(\omega ) $ is the arbitrary phase at the circular frequency $ \omega $. However, the selected method of $ |{RM_{est2}(\omega_c,\psi)|} $ and $ \chi_{est2}(\omega_c,\psi) $ estimation (see Eq. (\ref{eq24})) does not require the knowledge of the multiplier, since it gives the same results for any initial phase, identical in channels A and B.

\section{Results.}

\begin{figure}[h]                                                               %fig12
    %\imagei
    \includegraphics[width=84mm]{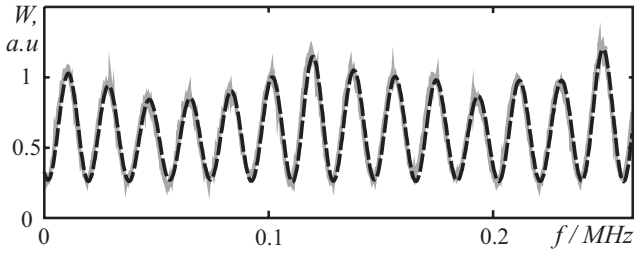}
    \caption{%
        Comparison of the source signal (solid grey line) with the model function in the central cross-section of the pulse (dashed black line). } %% (Поместить  и  в поле рисунка)no full stop at the end
    \label{fig12}
\end{figure}

\begin{figure}[h]                                                               %fig13
%   \imagei
    \includegraphics[width=84mm]{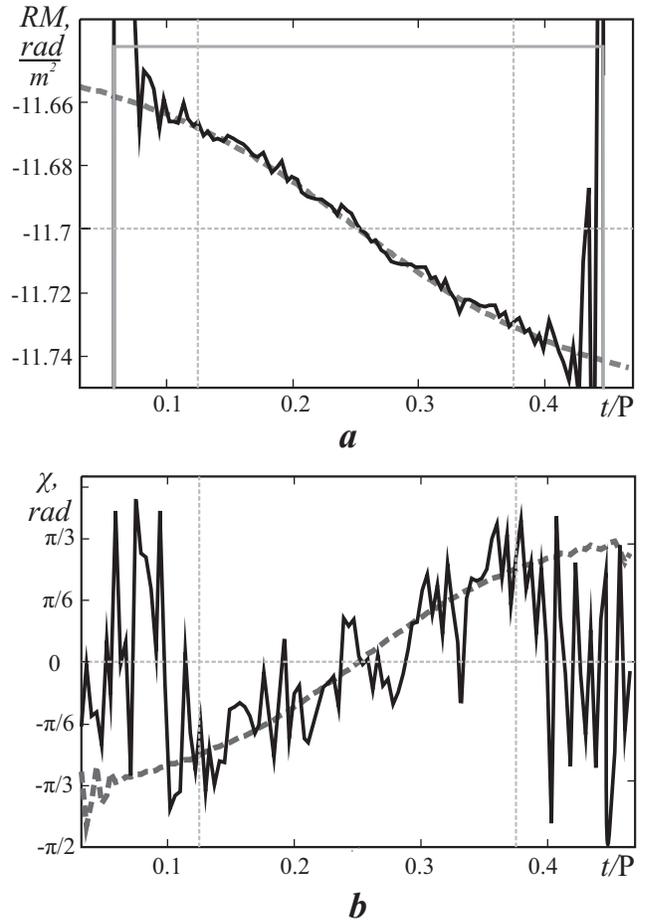}
    \caption{%
        Estimates of rotation measure $ RM_{est2}(\omega_c, \psi ) $ ($ a $)) and positional angle $ \chi_{est2}(\omega_c, \psi ) $ ($ b $)). In both cases the solid black curves denote the estimated values, for comparison the initial model parameters (denoted by the dashed line) are presented (see Fig. \ref{fig7} and Fig. \ref{fig3}e). At the top of $ a $) the solid gray line represents the preliminary estimation obtained from the polarization response  (see Fig. \ref{fig9}) by spectral analysis (see eq. (\ref{eq22})).} %% no full stop at the end
    \label{fig13}
\end{figure}

Below the results of the simulation are presented. Fig. \ref{fig12}-\ref{fig13}$ a $ show the comparison between the two $RM$ estimation methods. The first method uses a Fourier analysis of the linearly polarized response intensity. The second method gives the $RM$ estimates using the discrepancy minimizing (Eq. (\ref{eq24})). You can see that the curve, obtained by the second method, and inscribed in the registered response (see Fig. \ref{fig12}) simulates the properties of polarized radiation and its propagation medium significantly better. Even more clearly, this conclusion is presented in Fig. \ref{fig13}. Data shown in this figure indicate that using the proposed algorithm the pulsar magnetosphere can be resolved in depth along the pulse profile. Simple Fourier analysis of the polarization response does not allow for such a result. The position angle traverse at decameter wavelengths, in our opinion, can only be assessed by the proposed algorithm. It follows from the considerations above that the error in determining the rotation measure $ RM(\omega_с, \psi ) $ and the position angle traverse $ \chi(\omega_c, \psi) $ are interconnected: $ \delta RM(\omega_c, \psi) \cdot c^2/f_c^2 \sim \delta \chi(\omega_c, \psi ) $. Fig. \ref{fig13}$ a,b $ just show this relationship. Improvement of the estimates carried out for the rotation measure became possible due to a fuller use of the polarized pulsed radiation properties and a more accurate representation of the propagation medium model. In our case, the correlation property of polarized radiation in a narrow band in the source reference frame was used. This, in fact, allowed to use the overdetermined system of equations for estimating the polarization parameters, highly correlated in the analyzed frequency band. From a physical perspective, this means that the response was taken into account not only at one selected frequency as in the Fourier method, but the entire spectrum of registered intensity including zero component was analyzed. Precisely these opportunities allowed to improve the methodological precision in determining the rotation measure and the  position angle traverse in $ \sqrt{N\!q} $ times.

As shown above, the fluctuations of the order $ \delta \chi_{est 2}(\omega_c, \psi) = \pm \pi /3 $ in the position angle determination, which can be seen in Fig. \ref{fig13}$ b $) correspond to the fluctuations in the $RM$ determination of the order $ \delta RM_{est 2}(\omega_c, \psi) = \pm 0.0065 $ rad/m$^2$ at the central frequency of 23.7 MHz (Fig. \ref{fig13}$ a $). In our model, the value of $  RM_{mod}(\omega_c, a) = -11.7  $ rad/m$^2$ was pawned. This rotation measure value according to the \cite{atnfCite2014} \citep{2005AJ....129.1993M} is that of PSR B0809+74 $ RM_{B0809+74}(\omega_c, \psi)  = -11.7 \pm 1.3  $ rad/m$^2$. Hence the relative methodological precision of the rotation measure estimation that was reached by us for this pulsar at the frequency of 23.7 MHz is $ \delta RM_{est 2}(\omega_c, \psi)/RM_{B0809+74}(\omega_c, \psi) = \pm 5.6 \cdot 10^{-4} $. This is comparable with the relative accuracy achieved in the measurement of the dispersion measure for this pulsar in the frequency range of 17-31 MHz ($f_c$ = 24 MHz)  $ DM_{B0809+74}(f_c,\psi) = 5.755 \pm 0.003  $pc/sm$^3$ or in relative terms $ \delta DM/DM_{B0809+74}(f_c,\psi) = 5.2 \cdot 10^{-4}  $ \citep{2013MNRAS.431.3624Z}. Here, to estimate the values $ DM_{B0809+74}(f_c, \psi) $ we do not use the data from ATNF catalog because it shows this parameter with a large uncertainty.

\begin{figure}[h]                                                               %fig14
    %\imagei
    \includegraphics[width=80mm]{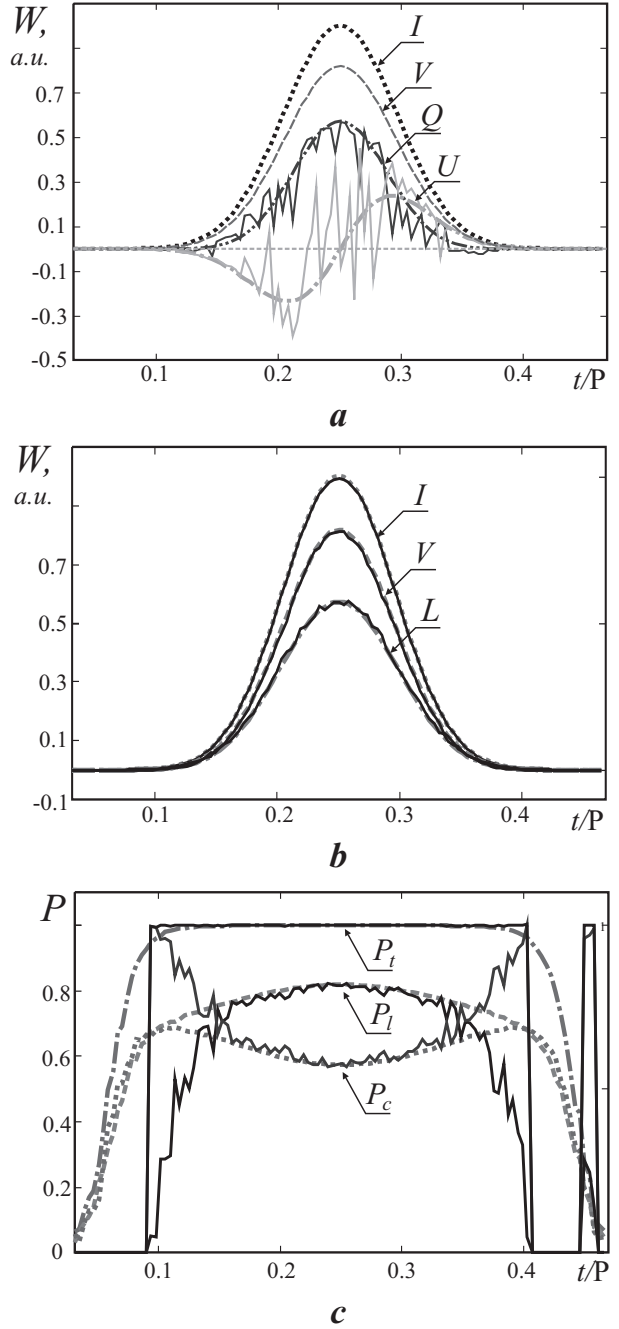}
    \caption{%
        Polarization parameters recovered in the source reference frame. $a$) is Stokes parameters, $b$) is the polarization invariants $I$, $L=\sqrt{Q^2 + U^2}$ and $V$, $c$) is the degrees of polarization $P_t$, $P_l$, $P_c$.} %% no full stop at the end
    \label{fig14}
\end{figure}

After the values $ |RM_{est2}(\omega_c, \psi )| $ and $ \chi_{est2}(\omega_c, \psi ) $  have become known, the further polarization parameters evaluation in the source reference frame is not a problem. It can be carried out, for example, by inverting equation (\ref{eq21}). An example of such evaluation in the framework of the developed model is shown in Fig. \ref{fig14}. Because we obviously took the ratio $S/N >$ 100, the main errors related to the polarization parameters determination in the source reference frame are defined   by errors in the values and  used in the inversion of the equations (\ref{eq21}) .

Thus, at this stage we were able to obtain estimates of the rotation measure and position angle  traverse from the signals recorded in the "free" space and restore all the polarization parameters of radiation in the source reference frame. Despite the fact that all the used above arguments are valid at any frequency range, it would be most reasonable to apply the algorithm in the decameter and meter bands. Relative registration band $ \Delta f/ f_c \sim 0.5 $ can be implemented most easily in these ranges where the decametric radio telescopes work. For accurate determination of $RM(f_{c})$ we should use the  relative frequency band not more then 0.1. Implementation of such registration parameters will allow, from our point of view, on the one hand to increase the aiming accuracy of rotation measure. On the other hand, and this is even more important it will allow to resolve the pulsar magnetosphere in depth \citep{2013IAUS..291..530U,UlyanovShevtsova2013} even at one pulse longitude. The latter circumstance is extremely important because it will be a fundamental breakthrough in observational studies not only of pulsars, but also the chemically peculiar stars with anomalously high magnetic fields. For these stars, we expect to find a connection with pulsars \citep{GopkaUlyanov2008,2008AIPC.1016..460G,2010AIPC.1269..454G,2010OAP....23...41G,2010AIPC.1269..448U,2008KPCB...24...36G, 2012OAP....25...35U}.

In the next part of the work we will show how one can estimate the original polarization parameters of pulsed radiation using considered model representations if the radiation is detected in LRF that is not in the "free" space but near the boundary between two media, fore example, air/ground or air/metal grid. Algorithms for the evaluation the initial polarization parameters that are optimal for different types of recording equipment will be considered.

\section{Conclusions}

In this paper, pulsed polarized signals are modeled. Polarization parameters of these signals change dynamically with a longitude of the mean pulse profile. On the basis of the eikonal equation the weakly anisotropic propagation medium is modeled. The parameters of this medium can change dynamically with time and depend on the pulse phase.

We proposed a new algorithm for estimating the parameters of the propagation medium. This algorithm at decameter wavelengths allows to reach the relative accuracy of $\sim  5 \cdot 10^{-4}$ in determining the rotation measure. Its use will allow not only significantly improve the existing estimates of the rotation measure in the direction to a number of pulsars, but also give the opportunity to resolve the upper pulsar magnetosphere in depth.

The use of this algorithm will result in the detection of fast fluctuations (from pulse to pulse or even inside the individual pulse) for both propagation medium parameters, and the polarization parameters of the pulsar radio emission. Ultimately, this will provide the better understanding of both the structure of the pulsar magnetosphere itself and the nature of its coherent radio emission.

\nocite{*}
\bibliographystyle{spr-mp-nameyear-cnd}
%\bibliography{myref}
\bibliography{biblio-Ulyanov}

\end{document}